\documentclass[11pt]{article}
\usepackage{setspace}
\usepackage{enumerate}
\usepackage{color}
\usepackage[pdftex]{graphicx}

\usepackage{subfigure}
\usepackage{amssymb,amsmath}
\usepackage{geometry}
\newcommand{\vect}[1]{\boldsymbol{#1}}
\normalsize

\usepackage{enumerate}
\usepackage{subfigure}
\usepackage{verbatim}
\usepackage{natbib}
\usepackage{multirow}
\usepackage{graphicx}
\usepackage{amssymb,amsmath}
\RequirePackage[OT1]{fontenc}
\RequirePackage{amsthm,amsmath}

\theoremstyle{plain}
\newtheorem{thm}{Theorem}
\newtheorem{lem}{Lemma}
\DeclareMathOperator*{\argmin}{\arg\!\min}

\begin{document}

\title  {Semiparametric Sieve Maximum Likelihood Estimation Under Cure Model with Partly Interval Censored and Left Truncated Data for Application to Spontaneous Abortion Data}

\vspace{20pt}
\author{Yuan Wu$^{1}$, Christina D.~Chambers$^{2,3}$ and Ronghui Xu$^{3,4}$\\
\small  $^1$Department of Biostatistics and Bioinformatics, Duke University, Durham, North Carolina. \\
\small $^2$Department of Pediatrics, $^3$Department of Family Medicine and Public Health, \\
\small $^4$Department of Mathematics,
University of California, San Diego.\\
}

\date{}
\maketitle


\begin{abstract}

This work was motivated by  observational studies in pregnancy with spontaneous abortion (SAB) as outcome. Clearly some women experience the SAB event but the rest do not. In addition, the data are left truncated due to the way pregnant women are recruited into these studies. For those women who do experience SAB, their exact event times are sometimes unknown. Finally, a small percentage of the women are lost to follow-up during their pregnancy. All these give rise to data that are left truncated, partly interval and right-censored, and with a clearly defined cured portion. We consider the non-mixture Cox regression cure rate model and adopt the semiparametric spline-based sieve maximum likelihood approach to analyze such data. Using modern empirical process theory
 we show that both the parametric and the nonparametric parts of the sieve estimator are consistent, and we establish the asymptotic normality for both parts. Simulation studies are conducted to establish the finite sample performance. Finally, we apply our method to a database of observational studies on spontaneous abortion.

\end{abstract}


\section{Introduction}
\label{s:intro}

Our work was motivated by research work carried out at
the Organization of Teratology Information Specialists (OTIS), which is a North American network of university or hospital based teratology services that counsel between 70,000 and 100,000 pregnant women every year.
Research subjects are enrolled from the Teratology Information Services and through other methods of recruitment, where the mothers and their babies are followed over time.
Recently it has been of interest to assess the effects of medication and vaccine exposures on spontaneous abortion (SAB).
By definition SAB occurs before 20 completed weeks of gestation; any pregnancy loss after that is called still birth.  Ultimately we would like to know if an exposure modifies the risk of SAB for a woman, which may be increased or decreased.
It is known that in the population
 for clinically recognized pregnancies the rate of SAB is  about 12\%.
 On the other hand, in our database the empirical SAB rate is consistently lower  than 10\%.
This is due to the fact that women may enter a study any time before 20 weeks' gestation.
The fact that we do not observe the women from the start of their pregnancy is known as left truncation in survival analysis; it reflects the selection bias in that women who have early SAB events can be seen as less likely to be in our studies.
In addition, a substantial portion of the SAB events do not have an exact known date, rather a window during which it occurred is typically available. This is known as interval censoring in survival analysis.
Finally, the fact that the majority of the pregnant women are  free of SAB is considered ``cured" in the time-to-event context.

Like in other clinical studies
our data also have right-censoring due to loss to follow-up before 20 weeks of gestation.
The typical survival analysis models  assume that all subjects in the study population will eventually experience the event of interest,
at least if they are not lost to follow-up.  When this is not the case,
in the literature researchers have proposed mixture and non-mixture cure models to deal with the situation. Mixture cure models have parts for the cure rate and the hazard function of the uncured subjects separately. The most popular semiparametric mixture cure rate model adopts logistic regression for the cure rate and Cox regression for the hazard rate. For example, \cite{sy:taylor:00} proposed the estimation under this model for right-censored data, \cite{ma:10} proposed the estimation under the same model for  interval censored data, and \cite{lam:xue:05}  and \cite{hu:xiang:16} adopted the sieve approach to ease  computation for interval censored data. Mixture cure models might have easy interpretation for practitioners,
 but are  computationally complex. On the other hand,
  the non-mixture cure model is easier to  compute,  and
has  become popular  for analyzing population with a well defined cured portion. For right-censored data, \cite{chen:ib:99} proposed a semiparametric method for a non-mixture cure model based on the Cox regression,  
and \cite{zeng:06} further extended the Cox regression  to general transformation models. For interval censored data, \cite{liu:shen:09} proposed a semiparametric method 
under the non-mixture cure model and established consistency of their estimator, \cite{hu:xiang:13} adopted the sieve approach for the nonparametric part and besides consistency they also established the asymptotic normality for the parametric part of the model.

In practical data analysis using cure models, a predetermined follow-up time window is often used to identify the observed cured subjects, see for example, \cite{sy:taylor:00} and \cite{zeng:06}. The end point of the follow-up window is called cure threshold by \cite{zeng:06}, and it is assumed that most or all events will occur before the cure threshold. In some applications, the cure threshold may be naturally defined related to the  events of interest. For example, spontaneous abortion (SAB) mentioned earlier is only defined as pregnancy loss before  week 20 of gestation, and subjects without such events before week 20 week are clearly ``cured" for SAB. Therefore, the cure threshold is naturally defined as week 20 for SAB. In this way, the cure model is also a natural candidate to be used for analyzing this type of data.

The fact that the SAB data consist of both interval censored and exactly observed  event times is referred as partly interval censored and actually occurs very often in practice.  Another example of partly interval censored data is progression free survival (PFS) time  in clinical trials, because PFS time is defined as the smaller of death and progression times which are usually right-censored and  interval censored, respectively.  Intuitively the asymptotic results for the maximum likelihood estimation (MLE) under the Cox model with partly interval censored data  will be the same as those for the MLE  with right-censored data in terms of convergence rate, since for both partly interval censored and right-censored data the likelihood function will be dominated by the term with observed events.
However,  \cite{kim:pic:03}  pointed out that if the interval censored observations are naively ignored from the whole data set, both estimation bias and standard error will be enlarged. Hence, a method correctly addressing this type of complicated data set is  needed.
Unfortunately we have found no published work on cure rate model with partly interval censored data in the presence of left truncation, which is the case for the SAB data application that we will describe in more details  in Section 7.
We will consider the sieve approach which has shown efficiency in  computation for both  nonparametric and semiparametric survival analysis problems under smoothness assumptions, and has variance estimator readily available.
\cite{ramsay} has observed that closely related to the well-known B-splines, there are so-called M-splines and I-splines, where the M-splines  are the derivatives of the  I-splines. In the following we will use the B-spline form for theoretical developments, and the M-spline and I-spline form for simplicity of computing.

To our best knowledge, this work is the first attempt to provide an 
approach for analyzing the complex survival data that are partly interval censored, left truncated and with a  cured portion. The paper is organized as follows. Section 2 proposes the semiparametric sieve MLE for the non-mixture Cox model when data are left truncated, partly interval censored and with a cured portion.
Section 3 provides all the asymptotic results for both the parametric part and the nonparametric part including consistency and asymptotic normality results. The convergence rate is showed to be the optimal for the nonparametric MLE problem. The asymptotic normality for the nonparametric part is established for the smooth functional of the sieve estimator. Section 4 describes the computational method for the proposed sieve MLE. Section 5 finds the estimator for the variance of both the parametric and the nonparametric part.  Section 6 conducts simulation studies to verify the finite sample performance for the proposed method. Section 7 applies the proposed methodology to  analyze an observational data set on spontaneous abortion.  Section 8 summaries the theoretical and numerical results  and discusses how it performs when our target data structure is simplified as several types and mentions some potential future work. In Appendix we provide proofs for all theorems in this paper with necessary lemmas using modern empirical process theory. 

\section{Semiparametric sieve MLE }

Consider the non-mixture cure model proposed by \cite{chen:ib:99}, in which the survival function of the event time $T$ given covariates $Z=z\in\mathbb{R}^d$ is
$S(t|z)=\exp\left\{-e^{\tilde{\beta}'\tilde{z}}F(t)\right\}$,
where $\tilde{\beta}=(\varphi_0,\beta')'$ is a vector of regression parameter contains an intercept $\varphi_0$ and $d$-dimensional vector $\beta$, $\tilde{z} = (1,z')'$ and  $0\le F(t)\le1$ is a distribution function. Let $F(\tau)=1$ and in the following we focus on the case when $\tau<\infty$. Since the  survival function here does not decrease beyond $\tau$, there are no subjects with $T>\tau$ and the cure threshold  is naturally equal to $\tau$  \cite[]{zeng:06}. In addition, subjects who do not experience the event within the time window $[0,\tau]$ are cured.

Let $Q$ be the left truncation time on $[0,\tau_1]$ with $0<\tau_1\le \tau$. And let $[U,V]$ be the observation interval on $[Q,\tau]$, where $U$ and $V$ may be both equal to $\tau$. 
 We assume that $T$ and $[U,V]$ are independent given $Z$ and $Q$, and $T$ and $Q$ are independent given $Z$.
   Denote $\Delta_1=I_{[U< T\le V]}$ for interval censoring, $\Delta_2=I_{[T>V]}$ for right-censoring and $\Delta_3=I_{[T\le U]}$ for observed events.

  Write $\Lambda(t)=F(t)\exp(\varphi_0)$, which represents the  baseline cumulative hazard for the non-mixture cure model. Note that  $\Lambda$ and  $(F,\varphi_0)$ have a one-to-one correspondence and $0\le\Lambda(t)\le \exp(\varphi_0)$ since $F$ is a distribution function and has maximum one.  With left truncated data the baseline cumulative hazard $\Lambda(\cdot)$ may not be reliably estimated due to lack of observations at the left end.
In this paper we 
   will show that nonetheless 
   the  functional increase of the baseline cumulative hazard  from a ``non-zero" point can be still accurately estimated.
  We note that since $\exp(\varphi_0)=\Lambda(\tau)$,  it will be also hard to estimate $\varphi_0$ with left truncated data, and here we avoid this estimation by focusing on $\Lambda(\cdot)$ and its increment.

In the following we  rewrite    the above non-mixture cure model  as
\begin{equation}\label{curemodel}
S(t|z)=\exp\left\{-e^{\beta'z}\Lambda(t)\right\},
\end{equation}
where $0\le\Lambda(t)\le \exp(\varphi_0)$, which is different from the unbounded cumulative baseline hazard in the original Cox model without cured subjects.
  Let $X=(T,U,V,Q,Z,\Delta_1,\Delta_2,\Delta_3)$ be the random observation and 
let $\lambda(t)$ satisfy $ \Lambda(t)=\int_0^t\lambda(u)du$.  The log-likelihood of an i.i.d.~sample $x_i=(t_i,u_i,v_i,q_i,z_i,\delta_{1,i},\delta_{2,i},\delta_{3,i})$, $i=1, ..., n$, based on the cure model (\ref{curemodel}) is

\begin{equation}\label{log_like1}
\begin{split}
l_{n}(\beta,\lambda;\cdot)=&\sum_{i=1}^{n}\delta_{1,i}\log\left(\exp\left[-e^{\beta'z_i}\{\Lambda(u_i)-\Lambda(q_i)\}\right]
-\exp\left[-e^{\beta'z_i}\{\Lambda(v_i)-\Lambda(q_i)\}\right]\right)\\
&+\sum_{i=1}^{n}\delta_{2,i}\left[-e^{\beta'z_i}\{\Lambda(v_i)-\Lambda(q_i)\}\right]\\
&+\sum_{i=1}^{n}\delta_{3,i}\left[-e^{\beta'z_i}\{\Lambda(t_i)-\Lambda(q_i)\}+\beta'z_i+\log \lambda(t_i)\right],
\end{split}
\end{equation}
by omitting the additive terms that do not involve $(\beta,\lambda)$.

The optimization of the above log-likelihood can be very challenging, as the semiparametric MLE approach would discretize $\lambda$ into point masses at each distinct observed event time, and under the continuous distribution assumption the number of distinct observations  is comparable to the sample size. We will then have to maximize (\ref{log_like1}) with a very large number of parameters when the  sample size is large.
To ease the computational difficulties for these type of estimation problems, \cite{geman:hwang:82} proposed a sieve maximum likelihood estimation procedure. The main idea of the sieve method is maximize the likelihood with much fewer variables in a subclass that ``approximates" to the original function space.
In addition, \cite{huang:tech:08} established that the sieve method provides an easy way to compute the 
observed information matrix. 
In the following
%
the sieve maximum likelihood estimation is proposed for the non-mixture cure model with partly interval censored and left truncated data.

Let the B-spline basis functions of order $l$ be $\left\{B_j^l(t)\right\}_{j=1}^{p_n}$ with knot sequence $\{\xi_j\}_{j=1}^{p_n+l}$ satisfying
$$0=\xi_1=\cdots=\xi_l<\xi_{l+1}<\cdots<\xi_{p_n}<\xi_{p_n+1}=\cdots=\xi_{p_n+l}=\tau,$$
where $p_n=O(n^\kappa)$ for $\kappa<1$.
With $\left\{B_j^l(t)\right\}_{j=1}^{p_n}$, define $$\Psi_n=\left\{\lambda_n=\sum_{j=1}^{p_n}\alpha_jB_j^l: \alpha_j\ge 0 ~\text{for}~j=1,\cdots,p_n\right\}.$$ The requirement for all coefficients being positive will guarantee that $\Psi_n$ only contains nonnegative function for approximating the space of smooth hazard functions on $[0,\tau]$.

If  $\lambda$ is replaced by $\lambda_n$ in (\ref{log_like1})
we have the log likelihood function as
\begin{equation}\label{log_like2}
\begin{split}
l_{n}(\beta,\lambda_n;&\cdot)\\
=&\sum_{i=1}^{n}\delta_{1,i}\log\left(\exp\left[-e^{\beta'z_i}\left\{\int_{q_i}^{u_i}\sum_{j=1}^{p_n}\alpha_jB_j^l(t)dt\right\}\right]\right.\\
&\left.-\exp\left[-e^{\beta'z_i}\left\{\int_{q_i}^{v_i}\sum_{j=1}^{p_n}\alpha_jB_j^l(t)dt\right\}\right]\right)\\
&+\sum_{i=1}^{n}\delta_{2,i}\left[-e^{\beta'z_i}\left\{\int_{q_i}^{v_i}\sum_{j=1}^{p_n}\alpha_jB_j^l(t)dt\right\}\right]\\
&+\sum_{i=1}^{n}\delta_{3,i}\left[-e^{\beta'z_i}\left\{\int_{q_i}^{t_i}\sum_{j=1}^{p_n}\alpha_jB_j^l(t)dt\right\}+\beta'z_i+\log \left\{\sum_{j=1}^{p_n}\alpha_jB_j^l(t_i)\right\}\right].
\end{split}
\end{equation}
The sieve maximum likelihood estimation is obtained through maximizing the log-likelihood function (\ref{log_like2}) in terms of $(\beta,\lambda_n)$. Note that the sieve MLE could have good asymptotic properties if $\Psi_n$ ``approximates"  the space of nonnegative functions.

\section{Asymptotic properties}

In this section, we describe asymptotic properties
of the proposed sieve semiparametric MLE. Study of the asymptotic properties of the proposed sieve estimator needs empirical process theory  
and requires
some regularity conditions, regarding the event time, observation time, truncation time and covariates. The following conditions sufficiently guarantee the results
in the forthcoming theorems.

\begin{enumerate}[C1]
\item
Covariate variable $Z$ is bounded, that is, there exists a scalar $z_0$ such that $|Z|<z_0$. Here $|\cdot|$ denotes Euclidean norm.
\item
For the true cumulative hazard $\Lambda_0(\cdot)$ for $T$, let $\lambda_0(\cdot)$ satisfy $\Lambda_0(t)=\int_0^t\lambda_0(u)du$. Then $\lambda_0(\cdot)$ is continuously differentiable up to order $p$ on $[0,\tau]$.
\item
If $T$ is interval censored, then $V-U$ has a uniform positive lower bound.
\item
Let $w(u|q,z)$ be the survival function of $U$ at $u$ given $Q=q$ and $Z=z$, then $w(u|q,z)$ has a uniform positive lower bound for $0\le u<\tau$ independent of $q$ and $z$.
\item
The joint density of $(T,U,V,Q,Z)$ has a uniform positive lower bound and a uniform upper bound in the the support region of joint random variable.
\item
For some $\eta\in(0,1)$, $a'Var(Z|T,Q)a\le \eta a'E(ZZ'|T,Q)a$ for all $a\in\mathbb{R}^d$.
\end{enumerate}

\noindent {\bf Remark 1.}~Condition C2 implies that $\lambda_0(t)$ is bound on $[0,\tau]$ and hence the survival rate of $T$ at $\tau$ is not 0, which corresponds cure rate model; Condition C2 also implies that the first derivative of $\lambda_0(t)$  is bounded on $[0,\tau]$, which is necessary to apply the result of Example 19.10 in \cite{vander:98} in the proof of consistency. Condition C3 guarantees the interval censored term in the likelihood function to be bounded.  Condition C4 implies that the conditional survival function of $U$ has a positive lower bound, which is a reasonable assumption since a significant portion of subjects are cured at the threshold $\tau$ ($U=\tau$).
Condition C5 implies that the density functions of $T$, $U$ and $V$ all have positive lower bounds and hence the data structure is truly partly interval censored including significant portions of observed events, interval censored events and right-censored events.
Condition C6 will be used similarly as C13 and C14  in \cite{wellner:zhang:07}.

Before stating our main theorems, we define some notations.
For the knot sequence $\{\xi_j\}_{j=1}^{p_n+l}$ previously defined for $\Psi_n$ with $p_n=O(n^\kappa)$ for $\kappa<1$, further let $\max_j\Delta_j= \max_{j=l,\cdots,p_n}(\xi_{j+1}-\xi_j)$ and $\min_j\Delta_j= \min_{j=l,\cdots,p_n}(\xi_{j+1}-\xi_j)$.  Then, with $\{\xi_j\}_{j=1}^{p_n+l}$ we define
\begin{align*}
\mathfrak{F}_n=&\left\{\lambda_n=\sum_{j=1}^{p_n}\alpha_j B_j^l: a_0\le \alpha_j\le K\tau b_0~\text{for}~j=1,\cdots,p_n, \right.\\
&\left.\frac{|\alpha_{j+1}-\alpha_j|}{\max_j\Delta_j}\le K^2d_0~\text{for}~j=1,\cdots,p_n-1, \int_0^\tau\lambda_n(t)dt\le \tau b_0,\right.\\ &\left.\frac{\max_j\Delta_j}{\min_j\Delta_j}~\text{has a  upper bound independent of}~n\right\},
\end{align*}
where $a_0, b_0$ and $d_0$ satisfy $a_0\le \lambda_0(t)\le b_0$ and $|\lambda'_0(t)|\le d_0$ on $[0,\tau]$, $K$ is a large positive number for relaxing the constraints on $\mathfrak{F}_n$ in finite sample computing as discussed in Section~4.

Note $a_0, b_0$ and $d_0$ do exist given C2 and C5. Then $\mathfrak{F}_n\subset\Psi_n$. Note that $\Psi_n$ is a general space of positive spline functions, and for the theoretical developments some regularity conditions are necessary to form $\mathfrak{F}_n$.   We also let $\mathbb{B}$ be a compact set in $\mathbb{R}^d$ and includes $\beta_0$ in its interior, and let $\theta=(\beta,\lambda)$ 
with $\beta\in \mathbb{B}$ and $\lambda\in\mathfrak{F}_n$. Then $\theta\in\Omega_n=(\mathbb{B},\mathfrak{F}_n)$.
We denote $\hat{\theta}_n=\left(\hat{\beta},\hat{\lambda}_n\right)$ the maximizer of $l_n(\theta;)$ over $\Omega_n$.

 Define $\|\cdot\|_{L_2(\nu)}$ the norm associated with the joint probability measure $\nu(t,q)$ for $(T,Q)$  based on the fact that $T\ge Q$,  as
  $\|f\|_{L_2(\nu)}=\int_0^\tau\int_q^\tau f^2(t)d\nu(t,q)$. Then we could define the distance between $\theta_1=(\beta_1,\lambda_1)$ and $\theta_2=(\beta_2,\lambda_2)$ as
 $$ d(\theta_1,\theta_2)=\left(|\beta_1-\beta_2|^2+\|\lambda_1-\lambda_2\|_{L_2(\nu)}^2\right)^{1/2}.$$

For one single observation $x=(t,u,v,q,z,\delta_1,\delta_2,\delta_2)$ from the random observation $X$ and a general semiparametric variable $\theta=(\beta,\lambda)$, the likelihood (after removing terms unrelated to $\theta$) is given by
\begin{align*}
l(\theta;x)=&\delta_1\log\left(\exp\left[-e^{\beta^Tz}\{\Lambda(u)-\Lambda(q)\}\right]
-\exp\left[-e^{\beta^Tz}\{\Lambda(v)-\Lambda(q)\}\right]\right)\\
&+\delta_2\left[-e^{\beta^Tz}\{\Lambda(v)-\Lambda(q)\}\right]\\
&+\delta_3\left[-e^{\beta^Tz}\{\Lambda(t)-\Lambda(q)\}+\beta^Tz+\log \lambda(t)\right],
\end{align*}
with $\Lambda(t)=\int_0^t \lambda(u)du$. We denote $\mathbb{M}(\theta)=Pl(\theta;x)$ with $P$ being the true joint probability measure of $(T,U,V,Q,Z,\Delta_1,\Delta_2,\Delta_3)$, and
$\mathbb{M}_n(\theta)=\mathbb{P}_nl(\theta;x)$ with
$\mathbb{P}_nf=\frac{1}{n}\sum_{i=1}^n f(x_i)$  the empirical process indexed by $f(X)$. Let $c$ be a constant that might have different values from place to place in the theoretical development. In what follows, we first  show the consistency of the proposed estimator and establish the rate of convergence.

\begin{thm}\label{thm:1}
Suppose that C1--C6 hold, then $\hat{\theta}_n$ is a consistent estimator for $\theta_0$ and
$$d(\hat{\theta}_n,\theta_0)= O_p\left(n^{- \min \left\{p\kappa,(1-\kappa)/2\right\} } \right).$$
\end{thm}


\noindent {\bf Remark 2.}~This theorem implies that for $\kappa=1/(1+2p)$, $d(\hat{\theta}_n,\theta_0)=O_p\left(n^{- p/(1+2p) } \right)$, which is the optimal convergence rate for the nonparametric MLE when the true target functions are continuously differentiable up to order $p$.
For any fixed $q$ and $t$ with $0 < q \le \tau_1$ and $q<t\le \tau$. Lemma 4 in the supplemental material implies that $\int_q^t\left\{\hat{\lambda}_n(x)-\lambda_0(x)\right\}^2dx< cd\left(\hat{\theta}_n,\theta_0\right)$. Hence, the estimation for the baseline hazard at any ``non-zero" point is fine and the functional increase  $\Lambda_0(t)-\Lambda_0(q)$ can be consistently estimated by the proposed sieve MLE. This is similar to the theoretical result for the estimated baseline hazard based on left truncated interval censored data in \cite{kim:left:03}.

Now we present the asymptotic normality for the proposed estimator including the parametric part and the smooth functional of the nonparametric part.
Consider a parametric smooth submodel with parameter $(\beta, \lambda_{(s,h)})$ with $\lambda_{(s,h)}=\lambda+sh$, then $\lambda_{(0,h)}=\lambda$, $\left.\frac{\partial \lambda_{(s,h)}}{\partial s}\right|_{s=0}=h$ and $\left.\frac{\partial \rho\left(\lambda_{(s,h)}\right)}{\partial s}\right|_{s=0}=\rho(h)$ for the functional $\rho(\cdot)$. Let $\mathcal{H}$ be the class of functions $h$ defined by this equation. The score operator for $\lambda$ with $h$ is the directional derivative at $\lambda$ along $h$:
$$l_\lambda(\theta;x)[h]=\left.\frac{\partial}{\partial s}l(\beta,\lambda_{(s,h)};x)\right|_{s=0}\equiv f_h(\beta,\lambda;x).$$
And the two times directional derivative at $\lambda$ along $h_1$ and $h_2$ is
$$l_{\lambda,\lambda}(\theta;x)[h_1][h_2]=\left.\frac{\partial}{\partial s}f_{h_1}(\beta,\lambda_{(s,h_2)};x)\right|_{s=0}.$$ In addition, for $\mathbf{h}=(h_1,\cdots,h_d)'$  with $h_s\in\mathcal{H}$ for $s=1,\cdots,d$, let $l_{\lambda}(\theta;x)[\mathbf{h}]$ be the d-dimensional vector with its $s$th element  $l_{\lambda}(\theta;x)[h_s]$. For $\mathbf{h}_1=(h_{1,1},\cdots,h_{1,d})'$ and
$\mathbf{h}_2=(h_{2,1},\cdots,h_{2,d})'$, let $l_{\lambda,\lambda}(\theta;x)[\mathbf{h}_1][\mathbf{h}_2]$ be the $d\times d$ matrix with its $i$th row $j$th column element  $l_{\lambda,\lambda}(\theta;x)[h_{1,i}][h_{2,j}]$.

For the $d$-dimensional $\beta=(\beta_1, ...,\beta_d)'$, let $l_{\beta}(\theta;x)=\{l_{\beta_1}(\theta;x),\cdots,l_{\beta_d}(\theta;x)\}'$, where $l_{\beta_s}(\theta;x)$ is the partial derivative of $l(\theta;x)$ with respect to $\beta_s$, $s=1,\cdots,d$.
Denote $\rho_s(\theta,h)=\left\{l_{\beta_s}(\theta;x)-l_\lambda(\theta;x)[h]\right\}^2$ for $s=1,\cdots,d$.   If $h_s^{\ast}=\arg \min_{h\in\mathcal{H}}P\rho_s(\theta_0,h)$, then by Theorem 1 on page 70 in \cite{bickel:93} the efficient score for $\beta_0$ is $l_\beta(\theta_0;x)-l_\lambda(\theta_0;x)\left[\mathbf{h}^{\ast}\right]$ with $\mathbf{h}^{\ast}=\left(h_1^{\ast},\cdots,h_d^{\ast}\right)'$. Let $\rho(\theta,\mathbf{h})=\left\{l_\beta(\theta;x)-l_\lambda(\theta;x)[\mathbf{h}]\right\}^{\otimes2}$,  then the information matrix for $\beta_0$ is given by
$$I(\beta_0)=P\rho(\theta_0,\mathbf{h}^{\ast}).$$

\begin{thm}\label{thm:2}
Suppose that C1--C6 hold,
$$\sqrt{n}\left(\hat{\beta}_n-\beta_0\right)=n^{-1/2}I^{-1}(\beta_0)\sum_{i=1}^n l^{\ast}(\theta_0;x_i)+o_P(1),$$
where $l^{\ast}(\theta; x)=l_\beta(\theta_0;x)-l_\lambda(\theta_0;x)\left[\mathbf{h}^{\ast}\right]$. That is,  $\sqrt{n}\left(\hat{\beta}_n-\beta_0\right)\rightarrow_d N\left(0, I^{-1}(\beta_0)\right)$ by the central limit theorem.
\end{thm}

Since the convergence rate we established is slower than $1/\sqrt{n}$, the asymptotic normality is not easy to obtain for $\hat{\lambda}_n(\cdot)$, the nonparametric part of the sieve MLE. However it can still be shown that the asymptotic normality is available for its smooth functional $\rho(\hat{\theta}_n)=\int_q^t\hat{\lambda}_n$, which is the plug in estimator of $\Lambda_0(t)-\Lambda_0(q)$ for any fixed $q$ and $t$ with $0<q\le \tau_1$ and $q<t\le \tau$. Here $q>0$ is chosen due to the left truncation, when the parameter cannot be estimated efficiently on the region close to zero. This corresponds to the consistency result for the nonparametric part we discussed in Remark 2. 


  The asymptotic normality of $\rho(\hat{\theta}_n)$ is established using the idea used in \cite{shen:97} and \cite{chen:fan:06}.

 Let $w=(w_{\beta}', w_\lambda)'$ with $w_{\beta}\in \mathbb{R}^d$ and $w_\lambda$ be a bounded function, then the directional derivative along $w$ of $l(\theta;x)$ evaluated at $\theta_0$ is given by
 \begin{equation}{\label{chen_d}}
 \frac{dl(\theta_0+tw;x)}{dt}|_{t=0}=\frac{dl(\theta_0;x)}{d\theta}[w]=l_\beta(\theta_0;x)'w_\beta+l_\lambda(\theta_0;x)[w_\lambda],
  \end{equation}
  where $l_\lambda(\theta_0;x)[w_\lambda]$ is as previously defined.
  Based on the directional derivative, the Fisher information  inner product is defined as
 $\langle w,\tilde{w}\rangle = P\left\{\left(\frac{dl(\theta_0;x)}{d\theta}[w]\right)\left(\frac{dl(\theta_0;x)}{d\theta}[\tilde{w}]\right)\right\}$ and the Fisher information
  distance is given by $\|w\|^2=\langle w,w\rangle$.

Now for the directional derivative of $\rho(\theta)$ at $\theta_0$, the Riesz representation theorem implies that there exists $w^{\ast}=({w_{\beta}^{\ast}}',w_{\lambda}^{\ast})'$ such that for any $w$ as defined above
$$\langle w^{\ast},w\rangle = \frac{d\rho(\theta_0)}{d\theta}[w]$$ and
$$\|w^{\ast}\|=\left\|\frac{d\rho(\theta_0)}{d\theta}\right\|^2=\sup_{\|w\|=1}\left|\frac{d\rho(\theta_0)}{d\theta}[w]\right|^2.$$
\begin{thm}\label{thm:3}
Given that C1--C6 hold,
$$\sqrt{n}\left\{\rho(\hat{\theta}_n)-\rho(\theta_0)\right\}\rightarrow_d N\left(0, \left\|\frac{d\rho(\theta_0)}{d\theta}\right\|^2\right),$$
with the finite variance $\left\|\frac{d\rho(\theta_0)}{d\theta}\right\|^2$.
\end{thm}

\section{Computing the sieve MLE}
In theoretical part we denoted the sieve MLE $\hat{\theta}_n=(\hat{\beta},\hat{\lambda}_n)$ as the maximizer of $l_n(\beta,\lambda_n;\cdot)$ defined by (\ref{log_like2}) over $\Omega_n=(\mathbb{B},\mathfrak{F}_n)$. In finite sample computing, we consider to relax the conditions for the $\alpha_j$'s in $\mathfrak{F}_n$. First for the spline knot sequence as in \cite{zhang:hua:huang} and \cite{wu:zhang:12} for sample size of distinct observations $n_0$  we let $p_n=[n_0^{1/3}]$, the largest integer smaller than $n_0^{1/3}$, and position interior knots based on quantiles of the data distribution. It can be seen that $\frac{\max_j\Delta_j}{\min_j\Delta_j}$ is naturally bounded since C5 implies distinct observations will be ``approximately" equally distributed. In $\mathfrak{F}_n$ the condition $\frac{|\alpha_{j+1}-\alpha_j|}{\max_j\Delta_j}\le K^2d_0$ implies that the difference between two adjacent I-spline coefficients is not large compared to $\max_j\Delta_j$, which will hold if $a_0\le\alpha_j\le K\tau b_0$ for finite sample size and large $K$.
Hence, if we define
$$\mathfrak{F}_n'=\left\{\lambda_n=\sum_{j=1}^{p_n}\alpha_jB_j^l: a_0\le\alpha_j\le K\tau b_0, ~\text{for}~j=1,\cdots,p_n, \int_0^\tau\lambda_n(t)dt\le \tau b_0 \right\}$$
with the knot sequence we just mentioned, then for finite sample computing we could replace $\mathfrak{F}_n$ by $\mathfrak{F}_n'$ and find the maximizer $\hat{\theta}$ of (\ref{log_like2}) over $\Omega_n'=(\mathbb{B},\mathfrak{F}_n')$. From the compactness of $\mathbb{B}$, we simply let $|\beta|\le c_0$ in computing.

We observe that in (\ref{log_like2}) the integration of the B-spline basis functions are involved, which complicates the computing. As an alternative to B-spline based sieve estimation, monotone I-spline technique for sieve estimation was first introduced by \cite{ramsay}. In what follows we choose to adopt the monotone I-splines to approximate the cumulative hazard $\Lambda_0(\cdot)$.  Thus the integration of B-spline basis functions can be avoided.
We note that  \cite{joly:98} also applied a similar computational approach for estimating hazard and cumulative hazard functions in survival data with a penalty term in the likelihood, but with no theoretical results.

Let $I_j^l$ and $M_j^l$ be I-spline and M-spline basis functions, respectively, as defined by \cite{ramsay} and \cite{schumaker:81}, with $M_j^l(t)=\frac{dI_j^l(t)}{dt}$. \cite{wu:zhang:12} showed that $I_j^l(t)=\sum_{k=j+1}^{p_n+1}B_k^{l+1}(t)$ and $M_j^l(t)=\frac{l}{\xi_{j+l}-\xi_j}B_j^l(t)$, where $\xi_{j+l},\xi_j$ are two knots from the knot sequence $\{\xi_k\}_{k=1}^{p_n+l}$ associated with the according B-spline basis functions. Note that $I_j^l$ has degree $l$, while both $B_j^l$ and $M_j^l$ have degree $l-1$.

 Then we can show that $\Phi_{N,n}=\left\{\int_0\lambda_n: \lambda_n\in\mathfrak{F}_n'\right\}$ is equivalent to $\mathfrak{F}_{I,n}$  with the I-spline function space $\mathfrak{F}_{I,n}$ defined as
 $$\mathfrak{F}_{I,n}=\left\{\Lambda_n=\sum_{j=1}^{p_n}\eta_jI_j^l: \sum_{j=1}^{p_n}\eta_j\le \tau b_0, a_0\le\frac{l}{\xi_{j+l}-\xi_j}\eta_j\le K\tau b_0,   ~\text{for}~j=1,\cdots,p_n\right\}.$$
  Hence, the B-spline based estimation problem can be converted to a equivalent I-spline based estimation problem. As just discussed, for finite sample case with large $K$ we could further simplify $\mathfrak{F}_{I,n}$ as
\begin{equation}\label{I_set}
\Phi_{I,n}=\left\{\Lambda_n=\sum_{j=1}^{p_n}\eta_jI_j^l: \sum_{j=1}^{p_n}\eta_j\le \tau b_0, \eta_j\ge m_j,   ~\text{for}~j=1,\cdots,p_n\right\},
\end{equation}
with each small positive number  $m_j=\frac{\xi_{j+l}-\xi_j}{l}a_0$.

Now we write the likelihood with I-spline basis functions as
\begin{equation}\label{log_like3}
\begin{split}
\tilde{l}_{n}(\beta,\Lambda_n;&\cdot)\\
=&\sum_{i=1}^{n}\delta_{1,i}\log\left(\exp\left[-e^{\beta'z_i}\left\{\sum_{j=1}^{p_n}\eta_jI_j^l(u_i)
-\sum_{j=1}^{p_n}\eta_jI_j^l(q_i)\right\}\right]\right.\\
&~~~~\left.-\exp\left[-e^{\beta'z_i}\left\{\sum_{j=1}^{p_n}\eta_jI_j^l(v_i)-\sum_{j=1}^{p_n}\eta_jI_j^l(q_i)\right\}\right]\right)\\
&+\sum_{i=1}^{n}\delta_{2,i}\left[-e^{\beta'z_i}\left\{\sum_{j=1}^{p_n}\eta_jI_j^l(v_i)
-\sum_{j=1}^{p_n}\eta_jI_j^l(q_i)\right\}\right]\\
&+\sum_{i=1}^{n}\delta_{3,i}\left[-e^{\beta'z_i}\left\{\sum_{j=1}^{p_n}\eta_jI_j^l(t_i)
-\sum_{j=1}^{p_n}\eta_jI_j^l(q_i)\right\}
+\beta'z_i+\log \left\{\sum_{j=1}^{p_n}\eta_jM_j^l(t_i)\right\}\right].
\end{split}
\end{equation}
 In practice for the finite sample I-spline based computing,
 we need to find the maximizer ${\tilde{\zeta}}_n=\left({\tilde{\beta}},{\tilde{\Lambda}}_n\right)\in\tilde{\Omega}_n=(\mathbb{B},\Phi_{I,n})$ for $\tilde{l}_{n}(\beta,\Lambda_n;\cdot)$ as defined by (\ref{log_like3}) over $\tilde{\Omega}_n$. Then by the aforementioned equivalency, we have $\tilde{l}_{n}(\tilde{\zeta}_n;\cdot)=l_n(\hat{\theta}_n;\cdot)$.
Since the constraints in $\Phi_{I,n}$ given by (\ref{I_set}) is made by linear inequalities, the maximization of (\ref{log_like3})  over $\tilde{\Omega}_n$ can be efficiently implemented by the generalized gradient algorithm \cite[]{jam:04}, as done in \cite{zhang:hua:huang} and  \cite{wu:zhang:12}. More details about this algorithm can be found in these papers.

\section{Variance estimation}

In addition to the advantage in computing the MLE, it is also straightforward to obtain the consistent observed information matrix for $\beta$ based on the proposed sieve MLE approach.
Denote $\mathbf{B}^l=\left(B_1^l, \cdots, B_{p_n}^l\right)'$ as the vector of B-spline basis functions of order $l$, then  $$l_\lambda\left(\hat{\theta}_n;x\right)\left[\mathbf{B}^l\right]
=\left\{l_\lambda\left(\hat{\theta}_n;x\right)\left[B_1^l\right],\cdots,l_\lambda\left(\hat{\theta}_n;x\right)
\left[B_{p_n}^l\right]\right\}'.$$
Let $A_{11}=\mathbb{P}_n\left[\left\{l_\beta\left(\hat{\theta}_n;x\right)\right\}^{\otimes2}\right]$,
$A_{12}=\mathbb{P}_n\left[l_\beta\left(\hat{\theta}_n;x\right)\left\{l_\lambda\left(\hat{\theta}_n;x\right)
\left[\mathbf{B}^l\right]\right\}'\right]$, $A_{21}=A_{12}'$ and
$A_{22}=\mathbb{P}_n\left[\left\{l_\lambda\left(\hat{\theta}_n;x\right)\left[\mathbf{B}^l\right]\right\}^{\otimes2}\right]$. The observed information matrix is given by $$\hat{O}=A_{11}-A_{12}A_{22}^{-1}A_{21}.$$

\begin{thm}\label{thm:4}
Given that C1-C6 hold, $$\hat{O}\rightarrow_P I(\beta_0).$$
\end{thm}

Next, we  propose how to estimate the variance $\left\|\frac{d\rho(\theta_0)}{d\theta}\right\|^2$ for the plug-in estimator $\rho(\hat{\theta}_n)$ for $\Lambda_0(t)-\Lambda_0(q)$.  We  consider a similar method as for the observed information matrix for $\beta$. 
In what follows we adopt the idea described in \cite{cheng:14}. Let $\hat{\lambda}_n=\sum_{j=1}^{p_n}\hat{\alpha}_jB_j^l$ with $\hat{\theta}=(\hat{\beta},\hat{\lambda}_n)$.  By the construction of $A_{11}$, $A_{12}$, $A_{21}$ and $A_{22}$ above, we can treat $\tilde{O}=A_{22}-A_{21}A_{11}^{-1}A_{12}$ as the observed information matrix for the spline coefficient vector $\hat{\alpha}=(\hat{\alpha}_1,\cdots,\hat{\alpha}_{p_n})'$.  Since
$\rho(\hat{\theta}_n)=\int_q^t\hat{\lambda}_n(x)dx=\int_q^t\sum_{j=1}^{p_n}\hat{\alpha}_jB_j^l(x)dx$,
we have
\begin{align*}
 \frac{\partial \rho(\hat{\theta}_n)}{\partial \hat{\alpha}}&=\left\{\frac{\partial \rho(\hat{\theta}_n)}{\partial \hat{\alpha}_1},\cdots, \frac{\partial \rho(\hat{\theta}_n)}{\partial \hat{\alpha}_{p_n}}\right\}'\\
 &=\left\{\int_q^tB_1^l(x)dx,\cdots,\int_q^tB_{p_n}^l(x)dx\right\}'\equiv\tilde{\omega}.
 \end{align*}
 Hence, by delta method the variance for $\rho(\hat{\theta}_n)$ can be estimated by $\tilde{\omega}'\tilde{O}^{-1}\tilde{\omega}$.

\section{Simulation studies}


In simulation studies we let all spline basis functions have order $l=3$, that is, we use quadratic B-spline and M-spline basis functions, cubic I-spline basis functions throughout the simulation.  We choose sample size as 200 and 500 with 1000 repetitions. The knot sequence for splines is chosen as described in Section 4.


Let $\beta_0=(0.7, -0.5)$ and let covariate $Z=(Z_1,Z_2)$, where $Z_1$ follows standard normal distribution and $Z_2$ follows Bernoulli distribution with probability $0.5$ of $Z_2=1$.
We generate event time $T$ with three cumulative hazard functions $\Lambda_{0,1}(\cdot)$, $\Lambda_{0,2}(\cdot)$ and $\Lambda_{0,3}(\cdot)$ satisfying  $\Lambda_{0,1}(t)=e^{1.2}\frac{1-e^{-t}}{1-e^{-4}}$, $\Lambda_{0,2}(t)=\frac{1-e^{-t}}{1-e^{-4}}$ and $\Lambda_{0,3}(t)=e^{-1.2}\frac{1-e^{-t}}{1-e^{-4}}$ for $0\le t\le4$, and $\Lambda_{0,1}(t)=e^{1.2}$, $\Lambda_{0,2}(t)=1$ and $\Lambda_{0,3}(t)=e^{-1.2}$ for $t>4$. Hence, for all cases the cure threshold $\tau=4$. Note that $\Lambda_{0,1}(\cdot)$  represents the situation with an average observed cure rate of $0.135$  (small cure rate), $\Lambda_{0,2}(\cdot)$ with an average   cure rate of $0.448$ (medium cure rate) and $\Lambda_{0,3}(\cdot)$ with an average  cure rate of $0.755$ (large cure rate).
For all three cumulative hazard functions we generate left truncation time $Q$ and observation interval $[U,V]$ in two different ways: 1) $Q$ is generated from Uniform $[0,1]$, and $[U,V]$ is generated from Uniform $[1,4.5]$; 2)  $Q$ is generated from Uniform $[0,4]$, and $[U,V]$ is generated from Uniform $[Q,4.5]$.  For both cases $U$ and $V$ are set to  $4$ if they are $>4$, and $U=V-0.005$ if $V-U<0.005$.
For 1) the resulting truncation rates in the uncured subjects with $\Lambda_{0,1}(\cdot)$, $\Lambda_{0,2}(\cdot)$ and $\Lambda_{0,3}(\cdot)$ are $0.654$, $0.501$, and $0.421$, respectively, and the resulting censoring rates in the remaining subjects after truncation are $0.108$, $0.163$ and $0.195$, respectively; for 2) the corresponding truncation and censoring rates
are $0.894$, $0.831$, and $0.794$,  and
$0.258$, $0.323$ and $0.359$, respectively.
We refer to the above two settings as relatively light versus heavy truncation and censoring in the following.


We use the proposed sieve MLE method to estimate the parametric part $\beta$ and the smooth functionals of the nonparametric part $\Lambda_{0,k}(t)-\Lambda_{0,k}(q)$ for $k=1,2$, and 3.
Due to limitation of space we present here results for the small and large cure rates ($k=1$ and 3), while the results for $k=2$ are in-between of these two cases and are available from the authors.
 Table \ref{tab:1} 
 and \ref{tab:2} present the results for estimating $\beta$ for $n=$ 200 and 500, and
 Table \ref{tab:3} 
 and \ref{tab:4} present the estimation results for $\Lambda_{0,k}(t)-\Lambda_{0,k}(q)$ with $k=1$ and 3,  $q=1$ and $t=1.5, 2.5, 3.5$, respectively.
The tables include estimation bias, sample standard deviation (SD) and  average estimated standard error based on the proposed estimated information matrix introduced in Section 5 (SE), and  coverage probability of nominal $95\%$ confidence intervals based on the estimated standard error ($95\%$ CP).

\begin{table}[htbp]
\centering\caption{Estimation of the parametric part with small cure rate}\label{tab:1}
\begin{tabular}{ccccccc}
   \hline\hline
   \multicolumn{7}{c}{Light truncation and censoring}\\
   \hline\hline
              &    & True value & Estimate & SD & SE & $95\%$ CP \\
              \hline\hline
   size=200&    $\beta_{0,1}$        & 0.7     &  0.711& 0.112&0.117&96.5\%\\
      &$\beta_{0,2}$       & -0.5       & -0.503 &0.177&0.186 &96.0\%\\
         \hline\hline
   size=500&    $\beta_{0,1}$        & 0.7     &  0.709& 0.069&0.071&95.7\%\\
               &$\beta_{0,2}$       & -0.5       & -0.501 &0.112&0.114 &95.4\%\\
          \hline\hline
          \multicolumn{7}{c}{Heavy truncation and censoring}\\
    \hline\hline
               &    & True value & Estimate & SD & SE & $95\%$ CP \\
              \hline\hline
   size=200&    $\beta_{0,1}$        & 0.7     &  0.736& 0.153&0.163&96.4\%\\
      &$\beta_{0,2}$       & -0.5       & -0.525 &0.244&0.256 &94.7\%\\
          \hline\hline
    size=500&    $\beta_{0,1}$        & 0.7     &  0.716& 0.098&0.097&95.7\%\\
               &$\beta_{0,2}$       & -0.5       & -0.510 &0.149&0.155 &95.5\%\\
          \hline\hline
\end{tabular}
\end{table}

\begin{table}[htbp]
\centering\caption{Estimation of the parametric part with large cure rate}\label{tab:2}
\begin{tabular}{ccccccc}
   \hline\hline
   \multicolumn{7}{c}{Light truncation and censoring}\\
   \hline\hline
              &    & True value & Estimate & SD & SE & $95\%$ CP \\
                  \hline\hline
   size=200&    $\beta_{0,1}$        & 0.7     &  0.717& 0.209&0.221&96.7\%\\
      &$\beta_{0,2}$       & -0.5               & -0.527 &0.405&0.412 &96.5\%\\
              \hline\hline
   size=500&    $\beta_{0,1}$        & 0.7     &  0.710& 0.129&0.129&95.3\%\\
      &$\beta_{0,2}$       & -0.5               & -0.524 &0.244&0.245 &95.0\%\\
          \hline\hline
    \multicolumn{7}{c}{Heavy truncation and censoring}\\
   \hline\hline
              &    & True value & Estimate & SD & SE & $95\%$ CP \\
                  \hline\hline
   size=200&    $\beta_{0,1}$        & 0.7     &  0.736& 0.381&0.468&96.8\%\\
      &$\beta_{0,2}$       & -0.5               & -0.619 &0.697&0.856 &99.0\%\\
              \hline\hline
   size=500&    $\beta_{0,1}$        & 0.7     &  0.720& 0.212&0.230&96.7\%\\
      &$\beta_{0,2}$       & -0.5               & -0.509 &0.399&0.427 &96.6\%\\
          \hline\hline
\end{tabular}
\end{table}

\begin{table}[htbp]
\centering\caption{Estimation of the cumulative hazard difference with small cure rate }\label{tab:3}
\begin{tabular}{ccccccc}
   \hline\hline
   \multicolumn{7}{c}{Light truncation and censoring}\\
   \hline\hline
              &    & True value & Estimate & SD & SE & $95\%$ CP \\
              \hline\hline
   size=200&    $\Lambda_{0,1}(1.5)-\Lambda_{0,1}(1)$        & 0.490       &  0.507           & 0.109&0.112&95.9\%\\
      &$\Lambda_{0,1}(2.5)-\Lambda_{0,1}(1)$       & 0.967                 & 0.980             &0.207&0.224 &96.8\%\\
       &$\Lambda_{0,1}(3.5)-\Lambda_{0,1}(1)$       & 1.142                & 1.142             &0.233&0.308 &98.3\%\\
                    \hline\hline
   size=500&    $\Lambda_{0,1}(1.5)-\Lambda_{0,1}(1)$        & 0.490    &  0.496              & 0.067&0.070&96.6\%\\
      &$\Lambda_{0,1}(2.5)-\Lambda_{0,1}(1)$       & 0.967               & 0.974               &0.126&0.133 &95.5\%\\
       &$\Lambda_{0,1}(3.5)-\Lambda_{0,1}(1)$       & 1.142               & 1.144             &0.148&0.171 &96.6\%\\
          \hline\hline
    \multicolumn{7}{c}{Heavy truncation and censoring}\\
    \hline\hline
                &    & True value & Estimate & SD & SE & $95\%$ CP \\
              \hline\hline
   size=200&    $\Lambda_{0,1}(1.5)-\Lambda_{0,1}(1)$        & 0.490       &  0.515           & 0.133&0.143&96.2\%\\
      &$\Lambda_{0,1}(2.5)-\Lambda_{0,1}(1)$       & 0.967                 & 1.020             &0.262&0.277 &96.4\%\\
       &$\Lambda_{0,1}(3.5)-\Lambda_{0,1}(1)$       & 1.142                & 1.210             &0.295&0.328 &97.7\%\\
                    \hline\hline
   size=500&    $\Lambda_{0,1}(1.5)-\Lambda_{0,1}(1)$        & 0.490    &  0.500              & 0.087&0.087&94.4\%\\
      &$\Lambda_{0,1}(2.5)-\Lambda_{0,1}(1)$       & 0.967               & 0.983               &0.155&0.160 &95.9\%\\
       &$\Lambda_{0,1}(3.5)-\Lambda_{0,1}(1)$       & 1.142               & 1.170             &0.178&0.185 &95.5\%\\
          \hline\hline
\end{tabular}
\end{table}

\begin{table}[htbp]
\centering\caption{Estimation of the cumulative hazard difference with large cure rate }\label{tab:4}
\begin{tabular}{ccccccc}
   \hline\hline
   \multicolumn{7}{c}{Light truncation and censoring}\\
   \hline\hline
              &    & True value & Estimate & SD & SE & $95\%$ CP \\
              \hline\hline
     size=200&    $\Lambda_{0,3}(1.5)-\Lambda_{0,3}(1)$           & 0.044        &0.046           &0.013& 0.015&96.9\%\\
      &$\Lambda_{0,3}(2.5)-\Lambda_{0,3}(1)$                     & 0.088      & 0.088        &0.028&0.032 &96.1\%\\
       &$\Lambda_{0,3}(3.5)-\Lambda_{0,3}(1)$                    & 0.104        & 0.104       &0.032&0.074 &99.1\%\\
       \hline\hline
   size=500&    $\Lambda_{0,3}(1.5)-\Lambda_{0,3}(1)$        & 0.044    &  0.046        & 0.008&0.009&96.1\%\\
      &$\Lambda_{0,3}(2.5)-\Lambda_{0,3}(1)$       & 0.088               & 0.088        &0.018&0.019 &95.4\%\\
       &$\Lambda_{0,3}(3.5)-\Lambda_{0,3}(1)$       & 0.104               & 0.104         &0.020&0.030 &98.6\%\\
          \hline\hline
   \multicolumn{7}{c}{Heavy truncation and censoring}\\
   \hline\hline
              &    & True value & Estimate & SD & SE & $95\%$ CP \\
              \hline\hline
     size=200&    $\Lambda_{0,3}(1.5)-\Lambda_{0,3}(1)$           & 0.044        &0.047           &0.021& 0.025&95.9\%\\
      &$\Lambda_{0,3}(2.5)-\Lambda_{0,3}(1)$                     & 0.088      & 0.094        &0.039&0.056 &97.6\%\\
       &$\Lambda_{0,3}(3.5)-\Lambda_{0,3}(1)$                    & 0.104        & 0.109       &0.044&0.070 &98.0\%\\
       \hline\hline
   size=500&    $\Lambda_{0,3}(1.5)-\Lambda_{0,3}(1)$        & 0.044    &  0.047        & 0.013&0.014&94.5\%\\
      &$\Lambda_{0,3}(2.5)-\Lambda_{0,3}(1)$       & 0.088               & 0.092        &0.025&0.029 &96.1\%\\
       &$\Lambda_{0,3}(3.5)-\Lambda_{0,3}(1)$       & 0.104               & 0.107         &0.029&0.033 &95.4\%\\
          \hline\hline
\end{tabular}
\end{table}

From the tables we can see that the simulation results for estimating both $\beta$ and the increments of the baseline cumulative function from $q$ to $t$ in general become more accurate when sample sizes increase or truncation and censoring becomes less severe, with smaller biases, more accurate standard errors compared to the sample standard deviations, and reduction in the variability of estimation. We also see that the   coverage probabilities of the confidence intervals are  generally acceptable.

\begin{figure}[htbp]
\begin{center}
			\scalebox{0.6}{\includegraphics[angle =0]{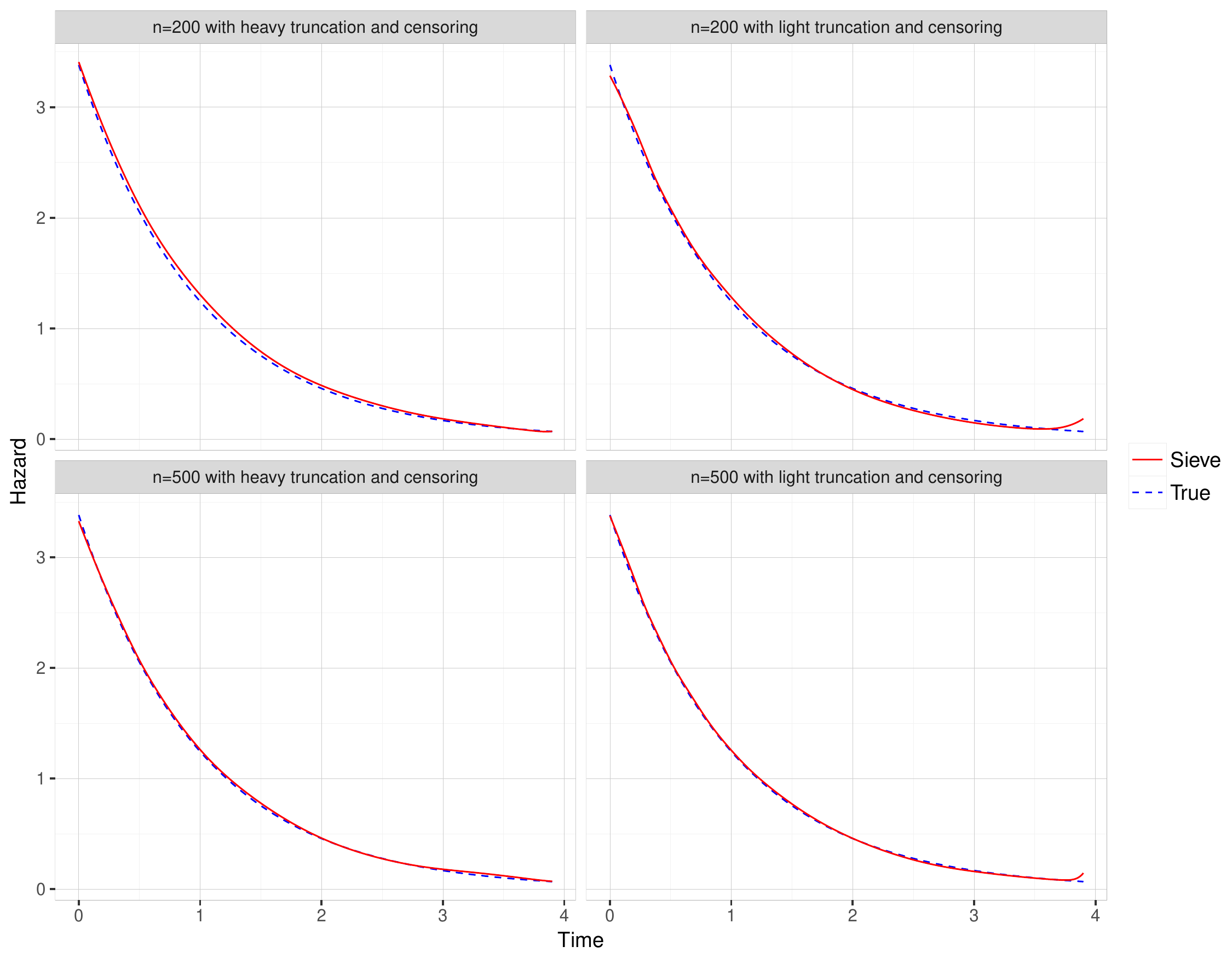}}
\caption{{ True baseline hazard function (True) and its sieve MLE (Sieve) with small cure rate.
 }}\label{fig:1}
\end{center}
\end{figure}

\begin{figure}[htbp]
\begin{center}
			\scalebox{0.6}{\includegraphics[angle =0]{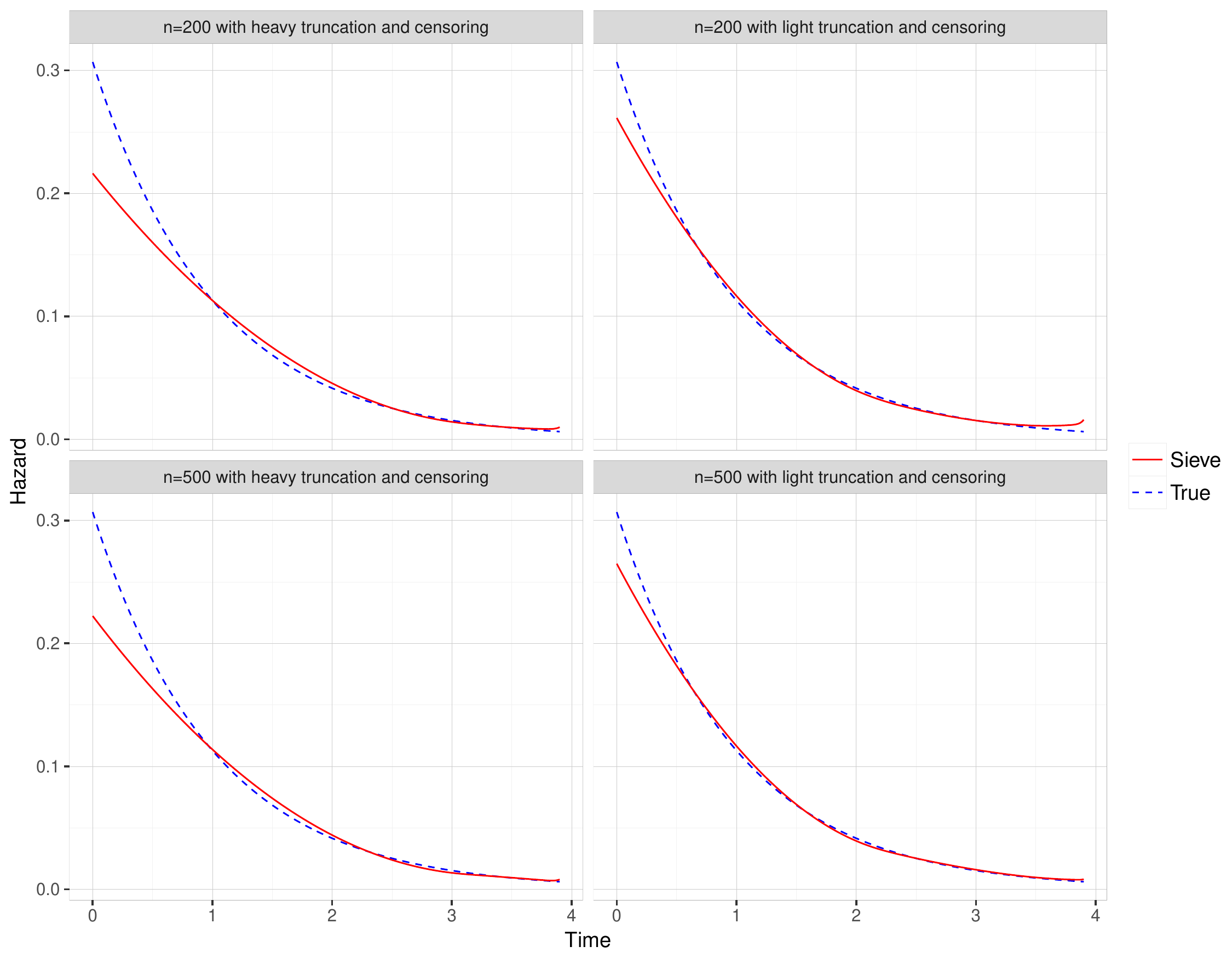}}
	\caption{{ True baseline hazard function (True) and its sieve MLE (Sieve) with large cure rate.
	 }}\label{fig:2}
\end{center}
\end{figure}

Finally, we also show the estimation results of the baseline hazard function $\lambda_{0,k}(\cdot)$ with $k=1$ and 3 on interval [0, 3.9] for  sample size 200 and 500, which are averaged over 1000 curves. Figure \ref{fig:1}
and \ref{fig:2} present the results for estimating $\lambda_{0,1}(\cdot)$ 
and $\lambda_{0,3}(\cdot)$, respectively.
We see that the estimation becomes more accurate for larger $n$. These figures also show that the estimation close to the right end point $\tau=4$ is not very accurate for light truncation and censoring and sample size 200, which is likely caused by the small number of the events around there.
We note that the estimation near $\tau=4$ seems to improve with heavier truncation and censoring, most likely because with heavier truncation more observations appear at later times (closer to $\tau=4$). 
In addition, it is important to note that the baseline hazard becomes noticeably underestimated close to time zero when the cure rate is larger and truncation and censoring is more severe. This underestimation phenomenon is likely caused by the reduced
risk sets due to left truncation and is consistent with our theoretical result that the estimation of the hazard function close to time zero is not reliable. However, we have noted earlier that the increments of the baseline cumulative function  can nonetheless be well estimated.

\section{Spontaneous abortion data analysis}

We apply the proposed sieve MLE method to an observational data set on spontaneous abortion from  the autoimmune disease in pregnancy database of the Organization of Teratology Information Specialists (OTIS) mentioned earlier. Our focus is to investigate the potential effect of  autoimmune disease medication on (spontaneous abortion) SAB, which is defined as any spontaneous pregnancy loss occurring before week 20 of gestation.

Our study sample includes pregnant women who entered a research
study between 2005 and 2012. It consists of 923 women who entered the study
before week 20 of their gestation. Since some women in the population may  experience the SAB event before having the chance to enter the study, we consider the study entry time  as left truncation time. Among the 923 subjects 56 women experienced the SAB event and the exact SAB time is known, 10 women also experienced the SAB event but only a time window including the incidence is available, 2 women were lost to follow-up before week 20, the rest of the women did not experience the SAB event.

In our proposed method, the lost to follow-up subjects and the observed cured subjects (subjects did not experience the SAB events before the cure threshold of week 20) are both treated as right-censored in the likelihood function under the non-mixture cure model, the same as in \cite{sy:taylor:00}. This way in the study sample we have 56 subjects with exact observed event times, 10 interval censored event times, and the rest are treated as right-censored. So the data set  is partly interval censored with left truncation, as women entered the research study any time during the first 20 weeks of gestation,  and also with a well defined cured portion. Since 10 interval censoring from all 66 women who experienced SAB is not an ignorable portion, the existing methods based on right-censoring is not applicable here.   Therefore the proposed sieve MLE method can be a good choice for the analysis.

For the primary comparison groups, among the 923 women 481 are pregnant and with certain autoimmune diseases which were treated with medications under investigation, 262  are women with the same specific autoimmune diseases but who were not treated with the medications under investigation, and the rest are healthy pregnant women without autoimmune diseases who were not treated with the medications.
We also include three important covariates: maternal age (range 18.6 - 47.1 years), prior therapeutic abortion (TAB; yes/no), and smoking (yes/no).
For the  analysis, as in the simulation studies we use quadratic B-spline and M-spline basis functions, and cubic I-spline basis functions. The knot sequence for the splines is chosen as described in Section 4.

Table \ref{tab:5} presents the estimation results for our study sample based on the proposed sive MLE approach.
According to the results from Table \ref{tab:5} we do not have statistical evidence to show that the autoimmune disease drugs have any significant  effects on the risk of SAB. We also see that older women have higher risk to experience the SAB events and smoking will increase the risk of the SAB.
Table \ref{tab:5} also shows the proposed sieve MLE for $\Lambda_0(t)$ and  $\Lambda_0(t)-\Lambda_0(q)$ with $t=17,18,19$ and $q=5$ (weeks). 
The standard errors of these estimates are consistent with our theoretical results and  imply that while the direct estimate for the baseline cumulative hazard function for the SAB occurring time has too much variability due to left truncation, the functional increase from a point not close to zero can still be reliably estimated.
\begin{table}[htbp]
\centering \caption{Estimation of covariate effects and cumulative baseline hazard using the spontaneous abortion data}\label{tab:5}
\begin{tabular}{crcc}
   \hline\hline
                      & Estimate  & SE & $p$-value \\
        Maternal age       & 0.079  & 0.025&0.002\\
     Prior Tab           &-0.358  & 0.436&0.411   \\
     Smoking              &  0.823 &0.364   &0.024 \\
     Healthy control        &-0.303 &0.479 &0.527 \\
          Diseased control       &0.236 &0.279& 0.398  \\
          \hline\hline
           $\Lambda_0(17)$        &  0.0173    &  0.020 &-\\
      $\Lambda_0(18)$       &  0.0174       & 0.020 &- \\
       $\Lambda_0(19)$       & 0.0174     & 0.020 &- \\
       $\Lambda_0(17)-\Lambda_0(5)$        & 0.0124 & 0.004   & -\\
      $\Lambda_0(18)-\Lambda_0(5)$       &  0.0125       & 0.004 &-\\
       $\Lambda_0(19)-\Lambda_0(5)$       & 0.0126       &0.004 & -\\
       \hline\hline
\end{tabular}
\end{table}

\begin{figure}
\begin{center}
			\scalebox{0.6}{\includegraphics[angle =0]{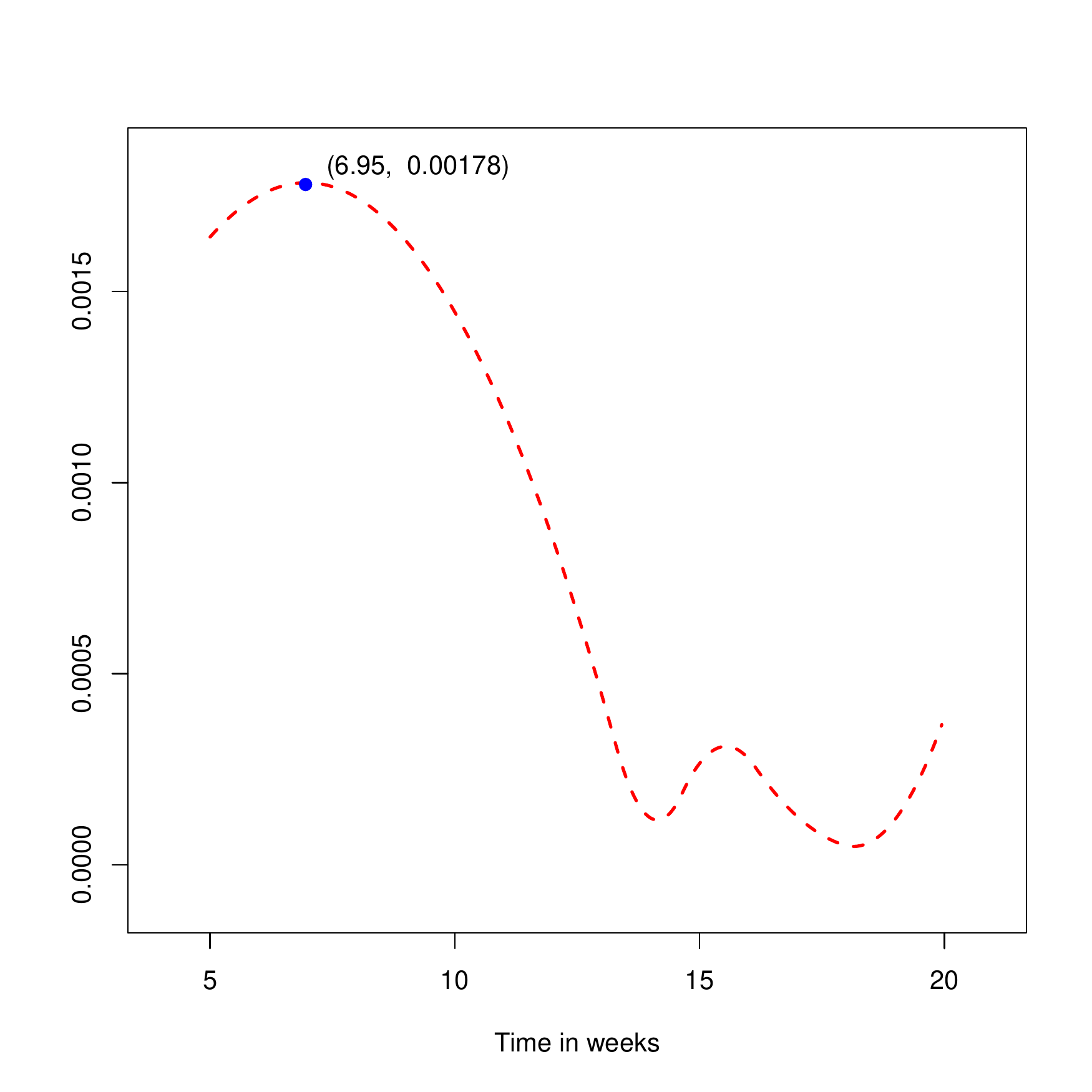}}
\caption{Estimated baseline hazard of spontaneous abortion}\label{fig:3}
\end{center}
\end{figure}

  Figure \ref{fig:3} shows the estimated baseline hazard function based on the proposed sieve MLE,
  and implies that the highest risk period  for women to experience the SAB events is between 5 and 10 weeks of gestation. This is consistent with existing scientific knowledge about spontaneous abortion.

\section{Concluding remarks}

In this paper we have proposed the semiparametric sieve MLE method to analyze  complex survival data that are partly interval censored, left truncated and with a cured portion. The proposed approach is motivated by a spontaneous abortion data application with this type of complicated structure, since no existing survival method is able to directly handle this type of survival data. Non-mixture cure model based on the Cox regression is used due to the relative simplicity of the likelihood computation. Using modern empirical process we have thoroughly studied the asymptotic properties for the proposed method: we have established that the proposed estimation is consistent with  optimal convergence rate for the nonparametric MLE problem; we have also established the asymptotic normality for both estimators of the parametric part and a functional of nonparametric part.  
In addition, we have provided closed form variance estimation for both the parametric and the nonparametric parts. In simulation studies we have showed that the finite sample performance of the proposed sieve MLE is satisfactory. Finally, the proposed model was successfully applied for analyzing the SAB data set.

    The proposed method is designed for relatively general survival data and usually applicable for simpler data structures. For different types of survival data, the proposed model may perform differently. For example, if partly interval censored data is replaced by right-censored only data, the proposed sieve MLE has the same asymptotic properties in terms of convergence rate and asymptotic normality as we mentioned in Section 1. However, if the data is purely interval censored, the estimation of hazard function will not be available based on the likelihood (since the third term in (\ref{log_like1}) disappears);
    separately by similar method as in \cite{zhang:hua:huang} it can be shown that the rate of estimation of cumulative hazard function will be slower than $\sqrt{n} $.
    In addition, if there is no left truncation,  the baseline cumulative hazard function itself can be  reliably estimated, as opposed to only its functional increases.

We have established that due to lack of data information around time zero for left truncated data, the nonparametric estimation around that region is not reliable. In the future we plan to tackle this issue and improve the estimation for the nonparametric part around time zero.
Another related potential work might be to replace the Cox model by the more general transformation model \cite[]{zeng:06} and developing the general semiparametric sieve MLE method.

\vspace{10pt}
\centerline{\Large{\sc Acknowledgements}}

\noindent
The research of Yuan Wu was supported in part by award number P01CA142538 from the National Cancer Institute.  The content is solely the responsibility of the authors and does not necessarily represent the official views of the National Institute of Health.

\vspace{10pt}
\centerline{\Large{\sc Appendix}}

\section*{Proofs for theorems}
{\em\underline {Proof of Theorem 1}}

\begin{enumerate}[(1)]
\item
We apply Theorem 5.7 in \cite{vander:98} to show the consistency. By the proof of Theorem 5.7 in \cite{vander:98}, we need to find a set including both $\theta_0$ and $\hat{\theta}_n$ as the ``$\Theta$" of Theorem 5.7 in \cite{vander:98}.
For this goal with enough small $a$ and enough large $b$ and $d$  first we define $\mathfrak{F}$ as
\begin{equation}\label{mathfrakf}
\begin{split}
\mathfrak{F}=&\left\{\lambda: 0<a\le \lambda(t)\le b<\infty ~\text{on}~[0,\tau], \right.\\
&\left. \lambda(t)=0~\text{for}~t<0~\text{or}~t>\tau, \lambda'(t)~\text{is continues on with}~|\lambda'(t)|\le d<\infty~\text{on}~[0,\tau]\right\}.
\end{split}
\end{equation}
And denote
\begin{equation}\label{omega}
\Omega=\left(\mathbb{B},\mathfrak{F}\right).
\end{equation}
Then Lemma \ref{lem:1} implies  $\theta_0\in\Omega$ and $\Omega_n\subset \Omega$, hence $\Omega$ include $\theta_0$ and $\hat{\theta}_n$. Then $\Omega$ is the ``$\Theta$". In what follows we complete the proof by verifying the conditions of Theorem 5.7 in \cite{vander:98}.
So $\mathfrak{F}_n\subset\mathfrak{F}$ when $a$ is small enough and $b$ and $d$ are large enough.
 Therefor $\theta_0\in\Omega$ and $\Omega_n\subset \Omega$, hence $\Omega$ include $\theta_0$ and $\hat{\theta}_n$. Then $\Omega$ is the ``$\Theta$". In what follows we complete the proof by verifying the conditions of Theorem 5.7 in \cite{vander:98}.

First, we verify $\sup_{\theta\in\Omega}\left|\mathbb{M}_n(\theta)-\mathbb{M}(\theta)\right|\rightarrow_p 0$.
Denote $\mathfrak{L}=\{l(\theta;x):\theta\in\Omega\}$, then
it suffices to show that $\mathfrak{L}$ is a $P$-Glivenko-Cantelli, since
$\sup_{\theta\in\Omega}\left|\mathbb{M}_n(\theta)-\mathbb{M}(\theta)\right|
=\sup_{l(\theta;X)\in\mathfrak{L}}\left|(\mathbb{P}_n-P)l(\theta;X)\right|\rightarrow_p 0$.

Since compact set $\mathbb{B}$ can be covered by $[c(1/\epsilon)^d]$ balls with radius $\epsilon$,  it can be shown that for any $\beta\in\mathbb{B}$ there exists a $\beta_{i_0}$ with ${i_0}\in\{1,2,\cdots,[c(1/\epsilon)^d]\}$ such that $|\beta-\beta_{i_0}|\le \epsilon$,
and hence $|\beta'z-\beta_{i_0}'z|\le c\epsilon$ for any $z$ by C1.

By Example 19.10 on Page 272 in \cite{vander:98}, we know that there exists $\parallel\cdot\parallel_\infty \epsilon$-brackets
$[\lambda_1^L, \lambda_1^R], \cdots, [\lambda_{[e^{c/\epsilon}]}^L, \lambda_{[e^{c/\epsilon}]}^R]$ to cover $\mathfrak{F}$. It is obvious that for any $\lambda\in\mathfrak{F}$ and $\Lambda(t)=\int_0^t\lambda(w)dw$ ,
there exists $[\lambda_{j_0}^L, \lambda_{j_0}^R]$ such that $\lambda_{j_0}^L(t)\le \lambda(t)\le \lambda_{j_0}^R(t)$ for some ${j_0}\in \{1,\cdots,[e^{c/\epsilon}]\}$, then
$\int_s^t \lambda_{j_0}^L(w)dw\le \int_s^t \lambda(w)dw\le  \int_s^t \lambda_{j_0}^R(w)dw$ for both $s$ and $t$ $\in [0,M]$.

Hence, it is easy to construct a set of brackets $\left[l_{i,j}^L, l_{i,j}^R\right]$ with $i=1, \cdots, [c(1/\epsilon)^d]$ and
$j=1,\cdots,[e^{c/\epsilon}]$ that for any $l(\theta;x)\in \mathfrak{L}$  with any observation $x=(t,u,v,q,z,\delta_1,\delta_2,\delta_3)$ we have $l_{i,j}^L\le l(\theta;x)\le l_{i,j}^R$, where
\begin{align*}
l_{i,j}^L=&\delta_1\log\left(\exp\left[-e^{\beta_i'z+c\epsilon}\left\{\int_q^u \lambda_{j}^R(w)dw\right\}\right]
-\exp\left[-e^{\beta_i'z-c\epsilon}\left\{\int_q^v \lambda_{j}^L(w)dw\right\}\right]\right)\\
&+\delta_2\left[-e^{\beta_i'z+c\epsilon}\left\{\int_q^v \lambda_{j}^R(u)du\right\}\right]\\
&+\delta_3\left[-e^{\beta_i'z+c\epsilon}\left\{\int_q^t \lambda_{j}^R(u)du\right\}+\beta_i'z-c\epsilon+\log \lambda_j^L(t)\right],
\end{align*}
\begin{align*}
l_{i,j}^R=&\delta_1\log\left(\exp\left[-e^{\beta_i'z-c\epsilon}\left\{\int_q^u \lambda_{j}^L(w)dw\right\}\right]
-\exp\left[-e^{\beta_i'z+c\epsilon}\left\{\int_q^v \lambda_{j}^R(w)dw\right\}\right]\right)\\
&+\delta_2\left[-e^{\beta_i'z-c\epsilon}\left\{\int_q^v \lambda_{j}^L(u)du\right\}\right]\\
&+\delta_3\left[-e^{\beta_i'z-c\epsilon}\left\{\int_q^t \lambda_{j}^L(u)du\right\}+\beta_i^Tz+c\epsilon+\log \lambda_j^R(t)\right].
\end{align*}
It can be seen that $\parallel l_{i,j}^R-l_{i,j}^L\parallel_\infty\le c\epsilon$ by C1, C2, C3, C5 and Taylor's expansion (by some algebra using the properties of $\mathfrak{F}$).
This leads to the conclusion that $N_{[~]}(\epsilon,\mathfrak{L},\parallel\cdot\parallel_\infty)\le c(1/\epsilon)^d e^{c/\epsilon}$.

Then by $N_{[~]}(\epsilon,\mathfrak{L},L_1(P))\le N_{[~]}(\epsilon,\mathfrak{L},\parallel\cdot\parallel_\infty)$, we have $N_{[~]}(\epsilon,\mathfrak{L},L_1(P))\le c(1/\epsilon)^d e^{c/\epsilon}$. Hence, $\mathfrak{L}$ is a $P$-Glivenko-Cantelli
  by Theorem 2.4.1 in \cite{van:wellner:96}.

Second, Lemma \ref{lem:2} establishes that for $\theta_0=(\beta_0,\lambda_0)$ and $\theta\in\Omega$
 $$\mathbb{M}(\theta_0)-\mathbb{M}(\theta)\ge cd(\theta,\theta_0)^2.$$

Finally, we verify $\mathbb{M}_n\left(\hat{\theta}_n\right)-\mathbb{M}_n(\theta_0)\ge -o_P(1)$.
Lemma \ref{lem:3} establishes that there exists an $\lambda_n$ in $\mathfrak{F}_n$ such that $\|\lambda_n-\lambda_0\|_\infty\le c\left(n^{-pv}\right)$, then also
$\left\|\int_q^u\{\lambda_n(t)-\lambda_0(t)\}dt\right\|_\infty\le c\left(n^{-pv}\right)$.
For $\theta_n=(\beta_0,\lambda_n)$, it can be seen that $\theta_n\in\Omega_n$. Then since $\hat{\theta}_n$ is the MLE over $\Omega_n$, we have $$\mathbb{M}_n\left(\hat{\theta}_n\right)-\mathbb{M}_n\left(\theta_n\right)\ge 0.$$ Hence,
\begin{align*}
\mathbb{M}_n\left(\hat{\theta}_n\right)-\mathbb{M}_n(\theta_0)
&=\mathbb{M}_n\left(\hat{\theta}_n\right)-\mathbb{M}_n\left(\theta_n\right)+\mathbb{M}_n\left(\theta_n\right)-\mathbb{M}_n(\theta_0)\\
&\ge \mathbb{M}_n\left(\theta_n\right)-\mathbb{M}_n(\theta_0)\\
&=\left(\mathbb{P}_n-P\right)\{l(\theta_n;x)-l(\theta_0;x)\}+P\{l(\theta_n;x)-l(\theta_0;x)\}
\end{align*}

By C1, C2, C3, the construction of $\mathfrak{F}_n$ and Taylor's expansion, we know that
$$P\{l(\theta_n;x)-l(\theta_0;x)\}^2\rightarrow 0~~\text{as}~~n\rightarrow\infty.$$
Then
\begin{align*}
\rho_P\{l(\theta_n;x)-l(\theta_0;x)\}&=\left(P\left[\{l(\theta_n;x)-l(\theta_0;x)\}-P\{l(\theta_n;x)-l(\theta_0;x)\}\right]^2\right)^{1/2}\\
&\le \left[P\{l(\theta_n;x)-l(\theta_0;x)\}^2\right]^{1/2}\rightarrow0~~\text{as}~~n\rightarrow\infty.
\end{align*}

By $N_{[~]}(\epsilon,\mathfrak{L},L_2(P))\le N_{[~]}(\epsilon,\mathfrak{L},\parallel\cdot\parallel_\infty)$, we have $N_{[~]}(\epsilon,\mathfrak{L},L_2(P))\le c(1/\epsilon)^d e^{c/\epsilon}$. Then
\begin{align*}
J_{[~]}\left(\delta,\mathfrak{L},L_2(P)\right)&=\int_0^\delta \sqrt{\log N_{[~]}\left(\epsilon,\mathfrak{L},L_2(P)\right)} d\epsilon\le\int_0^\delta\sqrt{\log\left\{c(1/\epsilon)^d e^{c/\epsilon}\right\}}d\epsilon\\
&\le\int_0^\delta\sqrt{c\left(\frac{1}{\epsilon}\right)}d\epsilon\le c\int_0^\delta \epsilon^{-1/2}d\epsilon=c\delta^{1/2} <\infty.\\
\end{align*}
So $\mathfrak{L}$ is a donsker by Theorem 19.5 in \cite{vander:98}. Then by Corollary 2.3.12 in \cite{van:wellner:96} we have
$$\left(\mathbb{P}_n-P\right)\{l(\theta_n;x)-l(\theta_0;x)\}=o_P\left(n^{-1/2}\right).$$ 

In addition, by Cauchy-Schwarz inequality as $n\rightarrow\infty$
\begin{align*}
|P\{l(\theta_n;x)-l(\theta_0;x)\}|\le P|l(\theta_n;x)-l(\theta_0;x)|\le c\left[P\{l(\theta_n;x)-l(\theta_0;x)\}^2\right]^{1/2}\rightarrow0.
\end{align*}

Then $P\{l(\theta_n;x)-l(\theta_0;x)\}>-o(1)$. Hence,
$$\mathbb{M}_n\left(\hat{\theta}_n\right)-\mathbb{M}_n(\theta_0)\ge o_P\left(n^{-1/2}\right)-o(1)=-o_p(1).$$
This completes the proof of $d\left(\hat{\theta}_n,\theta\right)\rightarrow_P0$.
\item
Next we establish the rate of convergence
 by verifying the conditions of Theorem 3.4.1 in \cite{van:wellner:96}. To apply this theorem, we denote $\mathbb{M}_n(\theta)$ as $\mathbb{M}(\theta)$ and denote $d_n(\theta,\theta_n)$ as $d(\theta,\theta_n)$. And we let the maximizer of $\mathbb{M}(\theta)$ or the true parameter as $\theta_0=(\beta_0,\lambda_0)$.

Let $\theta_n=(\beta_0,\lambda_n)$ with $\lambda_n\in\mathfrak{F}_n$ and $d(\theta_n,\theta_0)=d(\lambda_n,\lambda_0)\le c\left(n^{-p\kappa}\right)$. We verify that for every $n$ and arbitrary $\delta$ with $\delta>\delta_n=n^{-p\kappa}$,
$$\sup_{\delta/2<d(\theta,\theta_n)<\delta, \theta\in\Omega_n}(\mathbb{M}(\theta)-\mathbb{M}(\theta_n))\le -c \delta^2$$

Since $d(\theta,\theta_0)\ge d(\theta,\theta_n)-d(\theta_0,\theta_n)\ge \delta/2-c\left(n^{-p\kappa}\right)\ge c\delta$ for large $n$. By C1, C2, C3, C5 and the construction of $\mathfrak{F}_n$ we can show that $\mathbb{M}(\theta_0)-\mathbb{M}(\theta_n)\le cd(\theta_0,\theta_n)\le c\left(n^{-p\kappa}\right)$. Then by the result in the consistency development for large $n$,
\begin{align*}
\mathbb{M}(\theta)-\mathbb{M}(\theta_n)&=\mathbb{M}(\theta)-\mathbb{M}(\theta_0)+\mathbb{M}(\theta_0)-\mathbb{M}(\theta_n)\\
&\le -c\delta^2+c\left(n^{-p\kappa}\right)\le -c\delta^2
\end{align*}

We shall find a function $\psi(\cdot)$ such that
$$E\left\{\sup_{\delta/2<\delta(\theta,\theta_n)<\delta,\theta\in\Omega_n}\sqrt{n}(\mathbb{M}_n-\mathbb{M})(\theta-\theta_n)\right\}\le c\frac{\psi(\delta)}{\sqrt{n}},$$
and $\delta\rightarrow\psi(\delta)/\delta^{\alpha}$ is decreasing for $\delta$, for some $\alpha<2$, and for $\gamma_n\le\delta_n^{-1}= n^{p\kappa}$, it satisfies
$$\gamma_n^2\psi(1/\gamma_n)\le c\sqrt{n}~~\text{for every}~~n.$$

For $\theta_n=(\beta_0,\lambda_n)$ as defined previously, let $$\mathfrak{L}_{n,\delta}=\left\{l(\theta;x)-l(\theta_n;x): \theta\in\Omega_n, \delta/2<d(\theta,\theta_n)<\delta\right\}.$$ First we evaluate the bracketing number of $\mathfrak{L}_{n,\delta}$. Let $$\mathbb{B}_\delta=\left\{\beta: \theta=\left(\beta,\lambda_n'\right)\in\Omega_n, \delta/2\le d(\theta_n,\theta)\le\delta\right\}$$  and $$\mathfrak{F}_{n,\delta}=\left\{\lambda_n': \theta=\left(\beta,\lambda_n'\right)\in\Omega_n,\delta/2\le d(\theta_n,\theta)\le \delta\right\}.$$

As $\mathbb{B}_\delta-\beta_0\in \mathbb{R}^d$, $\mathbb{B}_\delta-\beta_0$ can be covered by $\left[c\left(\delta/\epsilon\right)^d\right]$ balls with radius $\epsilon$, that is, for any $\beta\in \mathbb{B}_\delta-\beta_0$ there exists a $\beta_{i_0}$ with $i_0\in\{1,2,\cdots, \left[c\left(\delta/\epsilon\right)^d\right]\}$ such that $|\beta-\beta_{i_0}|\le \epsilon$ and hence $|\beta'z-\beta_{i_0}'z|\le c\epsilon$ by C1. This implies that
$\beta'z\in\left[\beta_{i_0}'z-c\epsilon, \beta_{i_0}'z+c\epsilon\right]$ for any $\beta \in \mathbb{B}_\delta-\beta_0$. Then for any $\beta\in \mathbb{B}_\delta$, there exists $s$ such that $\beta'z\in\left[\tilde{\beta}_{i_0}'z-c\epsilon, \tilde{\beta}_{i_0}'z+c\epsilon\right]$ with $\tilde{\beta}_{i_0}=\beta_{i_0}+\beta_0$.

Also by  Lemma 0.6 in \cite{wu:zhang:12}, there exists $\epsilon$-brackets $\left[D_{n,{j_0}}^L, D_{n,{j_0}}^R\right]$, ${j_0}=1,2,\cdots,\left[\left(\delta/\epsilon\right)^{cp_n}\right]$ to cover $\mathfrak{F}_{n,\delta}-\lambda_n$. Denote $F_{n,{j_0}}^L=D_{n,{j_0}}^L+\lambda_n$ and $F_{n,{j_0}}^R=D_{n,{j_0}}^R+\lambda_n$, then for any $\lambda_n'\in\mathfrak{F}_{n,\delta}$, there exists ${j_0}$, such that $D_{n,{j_0}}^L\le \lambda_n'-\lambda_n\le D_{n,{j_0}}^R$ or equivalently
$F_{n,{j_0}}^L\le \lambda_n'\le F_{n,{j_0}}^R$.

Hence, for any $l(\theta;x)\in\mathfrak{L}_{n,\delta}+l(\theta_n;x)$ there exist $l_{n,i,j}^L$ and $l_{n,i,j}^R$ with $i\in\{1,2,\cdots, \left[c\left(\delta/\epsilon\right)^d\right]\}$, $j\in \left\{1,2,\cdots,\left[\left(\delta/\epsilon\right)^{cp_n}\right]\right\}$ and $l_{n,i,j}^L\le l(\theta;x)\le l_{n,i,j}^R$, where
\begin{align*}
l_{n,i,j}^L=&\delta_1\log\left(\exp\left[-e^{\tilde{\beta}_i'z+c\epsilon}\left\{\int_q^uF_{n,j}^R(t)dt\right\}\right]
-\exp\left[-e^{\tilde{\beta}_i'z-c\epsilon}\left\{\int_q^vF_{n,j}^L(t)dt\right\}\right]\right)\\
&+\delta_2\left[-e^{\tilde{\beta}_i'z+c\epsilon}\left\{\int_q^vF_{n,j}^R(t)dt\right\}\right]\\
&+\delta_3\left[-e^{\tilde{\beta}_i'z+c\epsilon}\left\{\int_q^tF_{n,j}^R(t)dt\right\}+\tilde{\beta}_i'z-c\epsilon+\log F_{n,j}^L(t)\right]
\end{align*}
and
\begin{align*}
l_{n,i,j}^R=&\delta_1\log\left(\exp\left[-e^{\tilde{\beta}_i'z-c\epsilon}\left\{\int_q^uF_{n,j}^L(t)dt\right\}\right]
-\exp\left[-e^{\tilde{\beta}_i'z+c\epsilon}\left\{\int_q^vF_{n,j}^R(t)dt\right\}\right]\right)\\
&+\delta_2\left[-e^{\tilde{\beta}_i'z-c\epsilon}\left\{\int_q^vF_{n,j}^L(t)dt\right\}\right]\\
&+\delta_3\left[-e^{\tilde{\beta}_i'z-c\epsilon}\left\{\int_q^tF_{n,j}^L(t)dt\right\}+\tilde{\beta}_i'z+c\epsilon+\log F_{n,j}^R(t)\right]
\end{align*}
By some algebra using the properties of $\mathfrak{F}_n$, we can show that $\left|l_{n,i,j}^R-l_{n,i,j}^L\right|\le c\epsilon$. Hence, the $\epsilon$-bracketing number with $\|\cdot\|_\infty$ norm for $\mathfrak{L}_{n,\delta}+l(\theta_n;x)$ is $\left(\delta/\epsilon\right)^{cp_n}$. Then obviously $\mathfrak{L}_{n,\delta}$ also has the $\epsilon$-bracketing number $\left(\delta/\epsilon\right)^{cp_n}$ for $\|\cdot\|_\infty$ norm. Since $L_2$-norm is bounded by $\|\cdot\|_\infty$ norm, we have
$$N_{[~]}\left\{\epsilon, \mathfrak{L}_{n,\delta}, L_2(P)\right\}\le c N_{[~]}\left\{\epsilon, \mathfrak{L}_{n,\delta}, \|\cdot\|_\infty\right\}\le \left(\delta/\epsilon\right)^{cp_n}.$$
By C1, C2, C3, C5 and  some algebra using the properties of $\mathfrak{F}_n$, we can show that $P\{l(\theta;x)-l(\theta_n;x)\}^2\le cd(\theta,\theta_n)^2\le c\delta^2$ and $\mathfrak{L}_{n,\delta}$ is uniformly bounded.
In addition,
\begin{align*}
\tilde{J}_{[~]}\left\{\delta,\mathfrak{L}_{n,\delta},L_2(P)\right\}&=\int_0^\delta\sqrt{1+\log N_{[~]}\left\{\epsilon,\mathfrak{L}_{n,\delta}, L_2(P)\right\}}d\epsilon\\
&\le\int_0^\delta\sqrt{1+cp_n\log\left(\delta/\epsilon\right)}d\epsilon\le\int_0^\delta cp_n^{1/2}\left(\delta/\epsilon\right)^{1/2}d\epsilon=cp_n^{1/2}\delta
\end{align*}
Then Lemma 3.4.2 in \cite{van:wellner:96}
\begin{align*}
E_P\|\sqrt{n}(\mathbb{P}_n-P)\|_{\mathfrak{L}_{n,\delta}}\le c \tilde{J}_{[~]}\left\{\delta,\mathfrak{L}_{n,\delta},L_2(P)\right\}\left[1+\frac{\tilde{J}_{[~]}
\left\{\delta,\mathfrak{L}_{n,\delta},L_2(P)\right\}}{\delta^2\sqrt{n}}\right]
\le c\psi(\delta)/\sqrt{n},
\end{align*}
with $\psi(\delta)=p_n^{1/2}\delta+p_n/n^{1/2}$. It is easy to see that $\psi(\delta)/\delta$ is a decreasing function of $\delta$. Note that for $p_n=n^\kappa$,
\begin{align*}
n^{2p\kappa}\psi\left(1/n^{p\kappa}\right)=n^{2p\kappa}n^{\kappa/2}n^{-p\kappa}+n^{2p\kappa}n^\kappa n^{-1/2}=n^{1/2}(n^{p\kappa+\kappa/2-1/2}+n^{2p\kappa+\kappa-1})
\end{align*}
Therefore, if $p\kappa<(1-\kappa)/2$ then $n^{2p\kappa}\psi\left(1/n^{p\kappa}\right)\le 2n^{1/2}$. Moreover
$$n^{1-\kappa}\psi\left(1/n^{(1-\kappa)/2}\right)=n^{1-\kappa}n^{\kappa/2}n^{-(1-\kappa)/2}+n^{1-\kappa}n^\kappa/n^{1/2}=2n^{1/2}.$$
This implies if $r_n=n^{\min\left\{p\kappa,(1-\kappa)/2\right\}}$, then $r_n\le\delta_n^{-1}= n^{p\kappa}$ and $r_n^2\psi\left(1/r_n\right)\le cn^{1/2}$.

Since $\mathbb{M}_n\left(\hat{\theta}_n\right)-\mathbb{M}_n(\theta_n)\ge 0$ and $d\left(\hat{\theta}_n,\theta_n\right)\le d\left(\hat{\theta}_n,\theta_0\right)+d(\theta_0,\theta_n)\rightarrow0$ in probability. Therefore, it follows by Theorem 3.4.1 in \cite{van:wellner:96} that $r_nd\left(\hat{\theta}_n,\theta_n\right)=O_P(1)$. Hence, by $d(\theta_n,\theta_0)\le cn^{-p\kappa}$
\begin{align*}
r_nd\left(\hat{\theta}_n,\theta_0\right)\le r_nd\left(\hat{\theta}_n,\theta_n\right)+r_nd(\theta_n,\theta_0)
\le O_P(1)+r_ncn^{-p\kappa} = O_P(1)
\end{align*}
This establishes the convergence rate. ~~$\square$
\end{enumerate}

{\em\underline{Proof of Theorem 2}}

By  Theorem 8.1 in \cite{huang:tech:08}, we only need to verify the following three conditions:
\begin{enumerate}[B1]
\item
$\mathbb{P}_nl_\beta\left(\hat{\theta}_n; x\right)=o_P\left(n^{-1/2}\right)$ and $\mathbb{P}_nl_\lambda\left(\hat{\theta}_n;x\right)\left[\vect{h}^*\right]=o_P\left(n^{-1/2}\right)$,
\item
$\left(\mathbb{P}_n-P\right)\left\{l^*\left(\hat{\theta}_n;x\right)-l^*(\theta_0;x)\right\}=o_P\left(n^{-1/2}\right)$,
\item
$P\left\{l^*\left(\hat{\theta}_n;x\right)-l^*(\theta_0;x)\right\}
=-I(\beta_0)\left(\hat{\beta}_n-\beta_0\right)+o_P\left(\left|\hat{\beta}_n-\beta_0\right|\right)+o_P\left(n^{-1/2}\right)$.
\end{enumerate}

Without loss of generalization in the following arguments for the three conditions we assume $\beta$ is one dimensional, then $\vect{h}^*$ is also one dimensional and denoted as ${h}^*$ .
First we verify B1:

Since $\hat{\theta}_n$ is the sieve MLE, we know that $$\mathbb{P}_nl_\beta\left(\hat{\theta}_n;x\right)=0=O_P\left(n^{-1/2}\right).$$ By Jackson's Theorem on page 149 in \cite{deboor:01}, we could find $h_n^*\in\mathfrak{G}_n$
with $\mathfrak{G}_n=\left\{g_n:g_n(t)=\sum_{j=1}^{p_n}\beta_j B_j^l(t)\right\}$ being the arbitrary B-spline space on $[0,M]$ with $p_n=O\left(n^{-\kappa}\right)$ such that $\left\|h_n^*-h^*\right\|_\infty\le cn^{-p\kappa}$. We also know that $\mathbb{P}_nl_\lambda\left(\hat{\theta}_n;x\right)\left[h_n^*\right]=0$, which is the directional derivative for $l\left(\hat{\theta}_n;x\right)$ along $h_n^*$ at $\hat{\lambda}_n$ with $\hat{\theta}_n=\left(\hat{\beta},\hat{\lambda}_n\right)$. In addition, we have $$Pl\left\{\beta_0, \lambda_0+s\left(h^*-h_n^*\right);x\right\}\le Pl(\beta_0,\lambda_0;x)$$ for $s$ with small absolute value, then $Pl_\lambda(\theta_0;x)\left[h^*-h_n^*\right]=0$.
Then we can write $$\mathbb{P}_nl_\lambda\left(\hat{\theta}_n;x\right)\left[h^*\right]=I_{1,n}+I_{2,n},$$ where $$I_{1,n}=\left(\mathbb{P}_n-P\right)l_\lambda\left(\hat{\theta}_n;x\right)\left[h^*-h_n^*\right]$$ and $$I_{2,n}=P\left\{l_\lambda\left(\hat{\theta}_n;x\right)\left[h^*-h_n^*\right]-l_\lambda(\theta_0;x)\left[h^*-h_n^*\right]\right\}.$$

Let $\mathbb{A}_1=\left\{\theta:\theta\in\Omega_n, d(\theta,\theta_0)\le cn^{-p\kappa}\right\}$. Since for a fixed $\tilde{\theta}\in \mathbb{A}_1$ and any $\theta\in \mathbb{A}_1$ we have $d\left(\tilde{\theta},\theta\right)\le cn^{-p\kappa}$. Then by similar arguments as in convergence rate development, we can first show that
$$N_{[~]}\left\{\epsilon, \mathbb{A}_1, L_2(P)\right\}\le \left(\frac{cn^{-p\kappa}}{\epsilon}\right)^{cp_n}=\left(\frac{n^{-p\kappa}}{\epsilon}\right)^{cn^\kappa},$$
and therefore show that for $\mathbb{A}_2=\left\{l_\lambda(\theta;x):\theta\in \mathbb{A}_1\right\}$
$$N_{[~]}\left\{\epsilon, \mathbb{A}_2, L_2(P)\right\}\le \left(\frac{n^{-p\kappa}}{\epsilon}\right)^{cn^\kappa}.$$
In addition, it can be similarly shown that for $\mathbb{A}_3=\left\{h-h^*: h\in\mathfrak{G}_n, \left\|h-h^*\right\|\le cn^{-p\kappa}, p\ge 2\right\}$
$$N_{[~]}\left\{\epsilon, \mathbb{A}_3, L^2(P)\right\}\le \left(\frac{n^{-p\kappa}}{\epsilon}\right)^{cn^\kappa}.$$
Hence combining the bracketing numbers for $\mathbb{A}_2$ and $\mathbb{A}_3$,\\
 for $\mathbb{A}_4=\left\{l_\lambda(\theta;x)\left[h-h^*\right]: l_\lambda(\theta;x)\in \mathbb{A}_2, h-h^*\in \mathbb{A}_3\right\}$
$$N_{[~]}\left\{\epsilon, \mathbb{A}_4, L_2(P)\right\}\le \left(\frac{n^{-p\kappa}}{\epsilon}\right)^{cn^\kappa}$$

Then
\begin{align*}
J_{[~]}\left\{\delta,\mathbb{A}_4,L_2(P)\right\}&=\int_0^\delta\sqrt{\log N_{[~]}\left\{\epsilon, \mathbb{A}_4, L_2(P)\right\}}d\epsilon\\
&=\int_0^\delta\sqrt{cn^\kappa\log\left(\frac{n^{-p\kappa}}{\epsilon}\right)}d\epsilon\\
&\le cn^{\kappa/2}n^{-p\kappa/2}\int_0^\delta\epsilon^{-1/2}d\epsilon\le cn^{(\kappa-p\kappa)/2}\delta^{1/2}<c\delta^{1/2}<\infty
\end{align*}

Then by Theorem 19.5 in \cite{vander:98} we know $\mathbb{A}_4$ is a Donsker class. Since $$l_\lambda\left\{\hat{\theta}_n;x\right\}\left[h^*-h_n^*\right]\in \mathbb{A}_4$$ and  as $n\rightarrow\infty$ $$P\left\{l_\lambda\left(\hat{\theta}_n;x\right)\left[h^*-h_n^*\right]\right\}^2\le c\left\|h^*-h_n^*\right\|_\infty^2\rightarrow0$$ , then by Corollary 2.3.12 of \cite{van:wellner:96} we have $$I_{1,n}=o_P\left(n^{-1/2}\right).$$ By some algebra and Cauchy-Schwarz inequality, it can be shown that
\begin{align*}
I_{2,n}&=P\left\{l_\lambda\left(\hat{\theta}_n;x\right)\left[h^*-h_n^*\right]
-l_\lambda\left(\theta_0;x\right)\left[h^*-h_n^*\right]\right\}\\
&\le cd\left(\hat{\theta}_n,\theta_0\right)\left\|h^*-h_n^*\right\|_\infty=o_P\left(n^{-1/2}\right).
\end{align*}
Then, $\mathbb{P}_nl_\lambda\left(\hat{\theta}_n;x\right)\left[h^*\right]=I_{1,n}+I_{2,n}=o_P\left(n^{-1/2}\right)$. Hence, B1 holds.

Next, we verify B2:

Let $\mathbb{A}_5=\left\{l^*(\theta;x)-l^*(\theta_0;x): \theta\in\Omega_n, d(\theta,\theta_0)\le cn^{-p\kappa}\right\}$. Then by similar arguments as for verifying B1 we can show that
$$N_{[~]}\left\{\epsilon, \mathbb{A}_5, L_2(P)\right\}\le \left(\frac{n^{-p\kappa}}{\epsilon}\right)^{cn^\kappa}.$$
Using the preceding argument, we know $\mathbb{A}_5$ is a Donsker class. Since $l^*\left(\hat{\theta}_n;x\right)-l^*(\theta_0;x)\in \mathbb{A}_5$ by the convergence rate of $\hat{\theta}_n$, it can be shown
$$P\left\{l^*\left(\hat{\theta}_n;x\right)-l^*(\theta_0;x)\right\}^2\le d\left(\hat{\theta}_n, \theta_0\right)^2\rightarrow_P0~\text{as}~n\rightarrow\infty.$$
Hence, by Corollary 2.3.12 in \cite{van:wellner:96} B2 holds.

Finally, we verify B3:

\begin{align*}
P\left\{l^*\left(\hat{\theta}_n;x\right)-l^*(\theta_0;x)\right\}
=&P\left\{l^*\left(\hat{\beta}_n,\hat{\lambda}_n;x\right)-l^*(\beta_0,\lambda_0;x)\right\}\\
=&P\left\{l_\beta\left(\hat{\beta}_n,\hat{\lambda}_n;x\right)-l_\beta(\beta_0,\lambda_0;x)\right\}\\
&-P\left\{l_\lambda\left(\hat{\beta}_n,\hat{\lambda}_n;x\right)\left[h^*\right]-l_\lambda(\beta_0,\lambda_0;x)\left[h^*\right]\right.\}
\end{align*}

By bivariate Taylor's expansion and the convergence rate of $\hat{\theta}_n$ we have
\begin{align*}
P&\left\{l_\beta\left\{\hat{\beta}_n,\hat{\lambda}_n;x\right\}-l_\beta(\beta_0,\lambda_0;x)\right\}\\
&=P\left\{l_{\beta\beta}(\beta_0,\lambda_0;x)\left(\hat{\beta}_n-\beta_0\right)
+l_{\beta,\lambda}(\beta_0,\lambda_0;x)\left[\hat{\lambda}_n-\lambda_0\right](1-0)\right.\\
&~~~~~~~\left.+o_P\left(\left|\hat{\beta}_n-\beta_0\right|\right)+o_P\left(n^{-1/2}\right)\right\},
\end{align*}
and
\begin{align*}
P&\left\{l_\lambda\left(\hat{\beta}_n,\hat{\lambda}_n;x\right)\left[h^*\right]-l_\lambda(\beta_0,\lambda_0;x)\left[h^*\right]\right\}\\
&=P\left\{l_{\lambda,\beta}(\beta_0,\lambda_0;x)\left[h^*\right]\left(\hat{\beta}_n-\beta_0\right)
+l_{\lambda,\lambda}(\beta_0,\lambda_0;x)\left[h^*\right]\left[\hat{\lambda}_n-\lambda_0\right](1-0)\right.\\
&~~~~~~~\left.+o_P\left(\left|\hat{\beta}_n-\beta_0\right|\right)+o_P\left(n^{-1/2}\right)\right\}.
\end{align*}
By the definition of $h^*$ and Theorem 11.1 in \cite{vander:98},
\begin{align*}
P&\left\{l_{\beta,\lambda}(\beta_0,\lambda_0;x)\left[\hat{\lambda}_n-\lambda_0\right]
-l_{\lambda,\lambda}(\beta_0,\lambda_0;x)\left[h^*\right]\left[\hat{\lambda}_n-\lambda_0\right]\right\}\\
&=-P\left[\left\{l_\beta(\beta_0,\lambda_0;x)
-l_\lambda(\beta_0,\lambda_0;x)\left[h^*\right]\right\}\left\{l_\lambda(\beta_0,\lambda_0;x)\left[\hat{\lambda}_n
-\lambda_0\right]\right\}\right]=0.
\end{align*}
Still by Theorem 11.1 in \cite{vander:98},
$$P\left[\left\{l_\beta(\beta_0,\lambda_0;x)
-l_\lambda(\beta_0,\lambda_0;x)\left[h^*\right]\right\}\left\{l_\lambda(\beta_0,\lambda_0;x)\left[h^*\right]\right\}\right]=0,$$
then we have
\begin{align*}
I(\beta_0)=&P\left\{l^*(\beta_0,\lambda_0;x)^{\otimes2}\right\}\\
=&P\left[\left\{l_\beta(\beta_0,\lambda_0;x)
-l_\lambda(\beta_0,\lambda_0;x)\left[h^*\right]\right\}\left\{l_\beta(\beta_0,\lambda_0;x)\right\}\right]\\
=&-P\left\{l_{\beta\beta}(\beta_0,\lambda_0;x)-l_{\lambda,\beta}(\beta_0,\lambda_0;x)\left[h^*\right]\right\}.
\end{align*}
Now combining the preceding results, we have
\begin{align*}
P\left\{l^*\left(\hat{\theta}_n;x\right)-l^*(\theta_0;x)\right\}
=-I(\beta_0)\left(\hat{\beta}_n-\beta_0\right)+o_P\left(\left|\hat{\beta}_n-\beta_0\right|\right)+o_P\left(n^{-1/2}\right),
\end{align*}
which means B3 holds and completes the proof. ~~$\square$

{\em \underline{Sketch of proof of Theorem 3}}

To derive the asymptotic normality we need the following assumptions:

Assumption 1. Let $\delta_n>0$ satisfying $\|\hat{\theta}_n-\theta_0\|=O_p(\delta_n)$. Then there exists small $\epsilon>0$, such that $\delta_n^{3-\epsilon}=o\left(n^{-1}\right)$

Assumption 2. Let $\mathfrak{G}_n$ be the arbitrary B-spline space as set before, then
$\mathfrak{F}_n\subset\mathfrak{G}_n$. Then there exists $w_{\lambda,n}^*\in\mathfrak{G}_n$ such that for $w_n^*=\left({w_\beta^*}',w_{\lambda,n}^*\right)'$, we have $\|w_n^*-w^*\|=o\left(n^{-1/3}\right)$ hence  $\delta_n\|w_n^*-w^*\|=o\left(n^{-1/2}\right)$.

 Assumption 1 and 2 can be easily verified by convergence rate for $\hat{\theta}_n$ and Jackson's Theorem in \cite{deboor:01}.

Let $\frac{d^2l(\theta;x)}{d\theta ^2}[{w}][{w}]$ be the two times directional derivative of $l(\theta;x)$ at $\theta$ along $w$  as defined by (4) in main paper. Then C1, C2, C3 and C5 guarantee the boundedness of the terms from
the second directional derivative. And the following assumption follows.

Assumption 3. If $\|\tilde{\theta}-\theta_0\|\le c\delta_n$ and $\|w\|\le c\delta_n$ ,
$$\left|P\left(\frac{d^2 l(\tilde{\theta};x)}{d\theta ^2}[{w}][{w}]-\frac{d^2l(\theta_0;x)}{d\theta ^2}[{w}][{w}]\right)\right|\le c(n^{-1}).$$

Next, let
\begin{align*}
\mathfrak{M}=\left\{\frac{dl(\tilde{\theta};x)}{d\theta}[w]-\frac{dl(\theta_0;x)}{d\theta}[w]:
\tilde{\theta}\in (\mathbb{R}^d, \mathfrak{G}_n), \|\tilde{\theta}-\theta_0\|\le c(\delta_n),\right.\\
\left.~~~~~~~~~~~~~~~~~~ w\in(\mathbb{R}^d, \mathfrak{G}_n), \|w-w^*\|\le cn^{-1/3}\right\}.
\end{align*}
By evaluating the bracketing number for $\mathfrak{M}$, it can be shown that $\mathfrak{M}$ is a Donsker class. Then we can establish the following assumption by Corollary 2.3.12 of \cite{van:wellner:96}.

Assumption 4. If $\|\tilde{\theta}-\theta_0\|\le c\delta_n$,
$$(\mathbb{P}_n-P)\left(\frac{dl(\tilde{\theta};x)}{d\theta}[w_n^*]-\frac{dl(\theta_0;x)}{d\theta}[w_n^*]\right)=o(n^{-1/2}).$$

Then using Assumption 1-4 and following the proof of Theorem 1 in \cite{chen:fan:06} we can establish that
$$\sqrt{n}\{\rho(\hat{\theta}_n)-\rho(\theta_0)\}\rightarrow_d N\left(0, \left\|\frac{d\rho(\theta_0)}{d\theta}\right\|^2\right).$$
 Given C5, by Cauchy-Shwarz inequality and Lemma \ref{lem:4} we have
\begin{align*}
\left|\int_q^tw(x)dx\right|^2\le c\int_q^tw^2(x)dx\le c\left\|w\right\|_{L_2(\nu)}^2
\end{align*}
It is easy to see that $\left\|\frac{d\rho(\theta_0)}{d\theta}\right\|^2=\sup_{\|w\|=1}\left|\int_q^tw(x)dx\right|^2$.
Hence, $\left\|\frac{d\rho(\theta_0)}{d\theta}\right\|^2<\infty$. ~~$\square$

{\em \underline{Proof of Theorem 4}}

Let $\hat{\vect{h}}_n=\left(\hat{h}_{1,n},\cdots,\hat{h}_{d,n}\right)'$ with $\hat{h}_{s,n}=\argmin_{h\in\mathfrak{G}_n}\mathbb{P}_n\rho_s\left(\hat{\theta}_n,h\right)$. \cite{huang:tech:08} showed that $\hat{O}=\mathbb{P}_n\rho\left(\hat{\theta}_n,\hat{\vect{h}}_n\right)$. Hence, now we need to verify that
$$\mathbb{P}_n\rho\left(\hat{\theta}_n,\hat{\vect{h}}_n\right)\rightarrow_P I(\beta_0).$$

We first show that $\left\|\hat{\vect{h}}_n-\vect{h}^*\right\|_d\equiv\max_{1\le s\le d}\left\|\hat{h}_{s,n}-h_s^*\right\|_{L2(P)}\rightarrow_P0$.

Let $\mathbb{G}_1=\left\{\rho_s(\theta,h):\theta\in\Omega_n, d(\theta,\theta_0)\le cn^{-p\kappa}, h\in\mathfrak{G}_n, \|h-h_s^*\|_\infty\le cn^{-p\kappa}\right\}$. We see that $$N_{[~]}\left\{\epsilon, \mathbb{G}_1, L_1(P)\right\}<\left(\frac{n^{-p\kappa}}{\epsilon}\right)^{cn^\kappa}<\infty$$ so $\mathbb{G}_1$ is a Glivenko-Cantelli class. By the Jackson's Theorem on page 149 in \cite{deboor:01} there exists a $h_{s,n}^*\in\mathfrak{G}_n$ such that $\left\|h_{s,n}^*-h_s^*\right\|_\infty\le cn^{-p\kappa}$. Then Glivenko-Cantelli theorem and Dominated Convergence Theorem with regularity conditions
\begin{align*}
\mathbb{P}_n&\rho_s\left(\hat{\theta}_n,\hat{h}_{s,n}\right)-\mathbb{P}_n\rho_s\left(\hat{\theta}_n,h_s^*\right)\\
&\le\mathbb{P}_n\rho_s\left(\hat{\theta}_n,{h}_{s,n}^*\right)-\mathbb{P}_n\rho_s\left(\hat{\theta}_n,h_s^*\right)\\
&=\left(\mathbb{P}_n-P\right)\left\{\rho_s\left(\hat{\theta}_n,{h}_{s,n}^*\right)-\rho_s\left(\hat{\theta}_n,h_s^*\right)\right\}
+P\left\{\rho_s\left(\hat{\theta}_n,{h}_{s,n}^*\right)-\rho_s\left(\hat{\theta}_n,h_s^*\right)\right\}\\
&=o_P(1)
\end{align*}

 \cite{huang:tech:08} showed that $\hat{h}_{s,n}$ can be obtained from standard least-squares calculation and is a function of $\hat{\theta}_n$ , then $d(\hat{\theta}_n, \theta_0)=o_P(1)$ and regularity conditions imply that there exists function $\tilde{h}_s$ such that $\left\|\hat{h}_{s,n}-\tilde{h}_s\right\|_{L_2(P)}=o_P(1)$.

Let
$\mathbb{G}_2=\left\{\rho_s(\theta,h):\theta\in\Omega_n, d(\theta,\theta_0)\le cn^{-p\kappa}, h\in\mathfrak{G}_n, \|h-\tilde{h}_s\|_{L_2(P)}=o_P(1)\right\}$.
Then we find that $N_{[~]}\left\{\epsilon, \mathbb{G}_2,L_1(P)\right\}<\left(\frac{o(1)}{\epsilon}\right)^{cn^\kappa}<\infty$  so $\mathbb{G}_2$ is a Glivenko-Cantelli class.
It is obvious that ${\mathbb{G}_3}=\left\{\rho_s(\theta,h_s^*):\theta\in\Omega_n,d(\theta,\theta_0)\le cn^{-p\kappa}\right\}$ is also a Glivenko-Cantelli class.
We just showed that $\mathbb{P}_n\rho_s\left(\hat{\theta}_n,\hat{h}_{s,n}\right)\le \mathbb{P}_n\rho_s\left(\hat{\theta}_n,h_s^*\right)+o_P(1)$, then
\begin{align*}
\left(\mathbb{P}_n-P\right)&\rho_s\left(\hat{\theta}_n,\hat{h}_{s,n}\right)+P\rho_s\left(\hat{\theta}_n,\hat{h}_{s,n}\right)\\
&\le \left(\mathbb{P}_n-P\right)\rho_s\left(\hat{\theta}_n,h_s^*\right)+P\rho_s\left(\hat{\theta}_n,h_s^*\right)+o_P(1).
\end{align*}
Then by $\mathbb{G}_2$ and $\mathbb{G}_3$ both being Glivenko-Cantelli classes, Glivenko-Cantelli theorem results in
$$P\rho_s\left(\hat{\theta}_n,\hat{h}_{s,n}\right)\le P\rho_s\left(\hat{\theta}_n,h_s^*\right)+o_P(1).$$

Then by $\hat{\theta}_n\rightarrow_P\theta_0$ using Dominated Convergence Theorem
$$P\left\{\rho_s\left(\hat{\theta}_n,\hat{h}_{s,n}\right)-\rho_s\left(\theta_0,\hat{h}_{s,n}\right)\right\}=o(1)$$
and
$$P\left\{\rho_s\left(\hat{\theta}_n,h_s^*\right)-\rho_s\left(\theta_0,h_s^*\right)\right\}=o(1).$$
Then
$$P\rho_s\left(\theta_0,\hat{h}_{s,n}\right)-P\rho_s\left(\theta_0,h_s^*\right)\le o_P(1).$$
Then by $h_s^*$ is the minimum of $P\rho_s\left(\theta_0,h_s^*\right)$ and by the fact that for the continuous functional $h\rightarrow P\rho_s\left(\theta_0,h\right)$ the range is closed for a closed domain,  there exists for any $\epsilon>0$ a number $\eta>0$ such that
$P\rho_s(\theta_0,h)\ge P\rho_s\left(\theta_0,h_s^*\right)+\eta$ for any h with $\epsilon\le\left\|h-h_s^*\right\|_{L_2(P)}\le M$. Thus, for large $n$
\begin{align*}
\Pr&\left\{\left\|\hat{h}_{s,n}-h_s^*\right\|_{L_2(P)}\ge\epsilon\right\}\le\Pr\left\{P\rho_s\left(\theta_0,\hat{h}_{s,n}\right)\ge P\rho_s\left(\theta_0,h_s^*\right)+\eta\right\}\\
&\rightarrow0,
\end{align*}
 Then we know $\left\|\hat{h}_{s,n}-h_s^*\right\|_{l_2(P)}\rightarrow_P0$ for $1\le s\le d$. Hence, $\left\|\hat{\vect{h}}_n-\vect{h}^*\right\|_d\rightarrow_P0$.

Next, let $\mathbb{G}_4=\left\{\rho(\theta,\vect{h}): \theta\in\Omega_n, d(\theta,\theta_0)\le cn^{-p\kappa}, h_s \in\mathfrak{G}_n ~\text{for}~s=1,\cdots,d, \vect{h}=(h_1,\cdots,h_d)',\right.\\
\left. \left\|\vect{h}-\vect{h}^*\right\|=o(1)\right\}$. Since we can show $N_{[~]}\left\{\epsilon,\mathfrak{G}_n,L_1(P)\right\}\le \left(\frac{o(1)}{\epsilon}\right)^{cn^\kappa}$ so $\mathbb{G}_4$ is a Glivenko-Cantelli class. Also by both $\hat{\theta}_n$ and $\hat{h}_n$ are consistent, we have
\begin{align*}
\mathbb{P}_n\rho\left(\hat{\theta}_n,\hat{\vect{h}}_n\right)&=\left(\mathbb{P}_n-P\right)\rho\left(\hat{\theta}_n,\hat{\vect{h}}_n\right)
+P\rho\left(\hat{\theta}_n,\hat{\vect{h}}_n\right)\\
&\rightarrow_P P\rho\left(\theta_0,\vect{h}^*\right)=I(\beta_0).~~\square
\end{align*}

\section*{Technical lemmas}
\begin{lem}\label{lem:1}
 Given C2 and C5. Then both $\lambda_0$ and $\mathfrak{F}_n$ will belong to $\mathfrak{F}$, where $\mathfrak{F}$ is defined by (\ref{mathfrakf}) in the proof of Theorem 1.
\end{lem}

{\em\underline{Proof of Lemma \ref{lem:1}}}
By C2 and C5, it is obvious that $\lambda_0\in\mathfrak{F}$. By the property that the sum of all basses equal to 1 on $[0,\tau]$ for B-spline we
have $a_0\le \lambda_n(t)\le K\tau b_0$ on $[0,\tau]$. Next
\begin{align*}
|\lambda'_n(t)|&=\left|\sum_{j=1}^{p_n-1}\frac{(l-1)(\alpha_{j+1}-\alpha_j)}{\xi_{j+l}-\xi_{j+1}} B_{j+1}^{l-1}(t)\right|
\le \max_j\frac{|\alpha_{j+1}-\alpha_j|(l-1)}{\min_j\Delta_j}\\
&\le \max_j\frac{|\alpha_{j+1}-\alpha_j|(l-1)}{\max_j\Delta_j}\frac{\max_j\Delta_j}{\min_j\Delta_j}
\le cK^2d_0
\end{align*}
So $\mathfrak{F}_n\subset\mathfrak{F}$ when $a$ is small enough and $b$ and $d$ are large enough.

\begin{lem}\label{lem:2}
Given C3--C6, for $\theta_0=(\beta_0,\lambda_0)$ and $\theta\in\Omega$, where $\Omega$ is defined by (\ref{omega}) in the proof of Theorem 1. Then
 $$\mathbb{M}(\theta_0)-\mathbb{M}(\theta)\ge cd(\theta,\theta_0)^2.$$
\end{lem}

{\em\underline{Proof of Lemma \ref{lem:2}}}

Let $S_0(t|z), S(t|z)$ denote the survival function for $T=z$ conditional on $Z=z$ for true $\theta_0$ and for any $\theta\in\Omega$, respectively, and let $f_0(t|z)=-dS_0(t|z)/dt$ and $f(t|z)=-dS(t|z)/dt$. Then
 $L(\beta,\lambda)=\left\{\frac{S_{}(u|z)-S_{}(v|z)}{S_{}(q|z)}\right\}^{{\delta}_1}
\left\{\frac{S_{}(v|z)}{S_{}(q|z)}\right\}^{{\delta}_2}
\left\{\frac{f_{}(t|z)}{S_{}(q|z)}\right\}^{{\delta}_3}$, after removing terms unrelated to $(\beta,\lambda)$. And $L(\beta_0,\lambda_0)$ the true likelihood function  is defined similarly.

Let $dP/d\mu=\varrho$ for Lebesgue measure (dominating measure) $\mu$. It is easy to see $\varrho$ is closely related to $L(\beta_0,\lambda_0)$ since $P$ is the joint probability measure of $X$. Then C3, C5 and construction of $\varrho$ implies $\varrho$ has a positive upper bound and $\varrho/L(\beta_0,\lambda_0)$ has a positive lower bound. Hence by the proof of Lemma 5.35 in \cite{vander:98}
\begin{align*}
\mathbb{M}(\theta_0)-\mathbb{M}(\theta)&=P\log L(\beta_0,\lambda_0)-P\log L(\beta,\lambda)=P\log\frac{L(\beta_0,\lambda_0)}{L(\beta,\lambda)}\\
&\ge c\int \left(\sqrt{L(\beta_0,\lambda_0)}- \sqrt{L(\beta,\lambda)}\right)^2 d\mu\ge c\int \left(L(\beta_0,\lambda_0)- L(\beta,\lambda)\right)^2 \varrho d\mu\\
&= cP\left(L(\beta_0,\lambda_0)- L(\beta,\lambda)\right)^2.
\end{align*}
Since
\begin{align*}
P\left(L(\beta_0,\lambda_0)- L(\beta,\lambda)\right)^2&
=P\left[\delta_1\left\{\frac{S_0(U|Z)-S_0(V|Z)}{S_0(Q|Z)}-\frac{S(U|Z)-S(V|Z)}{S(Q|Z)}\right\}^2\right]\\
&+P\left[\delta_2\left\{\frac{S_0(V|Z)}{S_0(Q|Z)}-\frac{S(V|Z)}{S(Q|Z)}\right\}^2\right]\\
&+P\left[\delta_3\left\{\frac{f_0(T|Z)}{S_0(Q|Z)}-\frac{f(T|Z)}{S(Q|Z)}\right\}^2\right]
\end{align*}
Then
\begin{align*}
P\left(L(\beta_0,\lambda_0)- L(\beta,\lambda)\right)^2
&\ge P\left[\delta_3\left\{\frac{f_0(T|Z)}{S_0(Q|Z)}-\frac{f(T|Z)}{S(Q|Z)}\right\}^2\right]\\
&=E\left(E\left[\left.\delta_3\left\{\frac{f_0(T|Z)}{S_0(Q|Z)}-\frac{f(T|Z)}{S(Q|Z)}\right\}^2\right| T,Q,Z\right]\right)\\
&=E\left(\left\{\frac{f_0(T|Z)}{S_0(Q|Z)}-\frac{f(T|Z)}{S(Q|Z)}\right\}^2 E\left[\left.I_{[U\ge T]}\right|T,Q,Z\right]\right)\\
&=E\left(\left\{\frac{f_0(T|Z)}{S_0(Q|Z)}-\frac{f(T|Z)}{S(Q|Z)}\right\}^2 Pr\left(U\ge T|T,Q,Z\right)\right)\\
&=E\left(\left\{\frac{f_0(T|Z)}{S_0(Q|z)}-\frac{f(T|Z)}{S(Q|z)}\right\}^2\right.\\
&~~~~~~~~~\left.\cdot \int_{\ge T}\frac{f_{U|Z,Q}(u|Z,Q)f_{T|Z,Q}(T|Z,Q)f_{Q,Z}(Q,Z)}{f_{T|Z,Q}(T|Z,Q)f_{Q,Z}(Q,Z)}du\right)\\
&=E\left(\int_{\ge T} f_{U|Z,Q}(u|Z,Q) du\left\{\frac{f_0(T|Z)}{S_0(Q|Z)}-\frac{f(T|Z)}{S(Q|Z)}\right\}^2 \right)
\end{align*}
Then by C4,
\begin{equation}\label{formula1}
\begin{split}
P\left(L(\beta_0,\lambda_0)- L(\beta,\lambda)\right)^2
&\ge E\left[w(T_{-}|Z,Q)\left\{\frac{f_0(T|Z)}{S_0(Q|Z)}-\frac{f(T|Z)}{S(Q|Z)}\right\}^2\right]\\
&\ge E\left[w(\tau_{-}|Z,Q)\left\{\frac{f_0(T|Z)}{S_0(Q|Z)}-\frac{f(T|Z)}{S(Q|Z)}\right\}^2\right]\\
&\ge c E\left\{\frac{f_0(T|Z)}{S_0(Q|Z)}-\frac{f(T|Z)}{S(Q|Z)}\right\}^2
\end{split}
\end{equation}

And by C5, $f_0(t|z)$ has positive lower bounded, then
\begin{align*}
E&\left\{\frac{S_0(T|Z)}{S_0(Q|Z)}-\frac{S(T|Z)}{S(Q|Z)}\right\}^2=E_{T,Q,Z}
\left\{\frac{S_0(Q|Z)-S_0(T|Z)}{S_0(Q|Z)}-\frac{S(Q|Z)-S(T|Z)}{S(Q|Z)}\right\}^2\\
&= E_{T,Q,Z}\left[\int_Q^T\left\{\frac{f_0(u|Z)}{S_0(Q|Z)}-\frac{f(u|Z)}{S(Q|Z)}\right\}du\right]^2.
\end{align*}
Then by Cauchy-Schwarz inequality
\begin{align*}
E&\left\{\frac{S_0(T|Z)}{S_0(Q|Z)}-\frac{S(T|Z)}{S(Q|Z)}\right\}^2
\le cE_{T,Q,Z}\left(\int_Q^T\left\{\frac{f_0(u|Z)}{S_0(Q|Z)}-\frac{f(u|Z)}{S(Q|Z)}\right\}^2du\right)\\
&= c\left(\int_{z,q}\int_q^\tau\left[\int_q^t\left\{\frac{f_0(u|z)}{S_0(q|z)}-\frac{f(u|z)}{S(q|z)}\right\}^2 du\right] f_0(t|z)f_{Q|Z}(q|z)f_Z(z) dt  dq dz \right)\\
&\le c\left(\int_{z,q}\int_q^\tau\left[\int_q^t\left\{\frac{f_0(u|z)}{S_0(q|z)}-\frac{f(u|z)}{S(q|z)}\right\}^2 f_0(u|z) du\right] f_0(t|z)f_{Q|Z}(q|z)f_Z(z) dt  dq dz \right)\\
&\le c\left(\int_{z,q}\int_0^\tau\left[\int_q^\tau\left\{\frac{f_0(u|z)}{S_0(q|z)}-\frac{f(u|z)}{S(q|z)}\right\}^2 f_0(u|z) du \right]f_0(t|z)f_{Q|Z}(q|z)f_Z(z) dt  dq dz  \right)\\
&= c\left(\int_{z,q}\left[\int_q^\tau\left\{\frac{f_0(u|z)}{S_0(q|z)}-\frac{f(u|z)}{S(q|z)}\right\}^2 f_0(u|z) du\right]\left[\int_0^\tau f_0(t|z) dt\right] f_{Q|Z}(q|z)  f_Z(z) dq dz \right)\\
&\le c\left(\int_{z,q}\int_q^\tau\left\{\frac{f_0(t|z)}{S_0(q|z)}-\frac{f(t|z)}{S(q|z)}\right\}^2 f_0(t|z)f_{Q|Z}(q|z)f_Z(z) dt  dq dz\right)\\
&= cE_{T,Q,Z}\left\{\frac{f_0(T|Z)}{S_0(Q|Z)}-\frac{f(T|Z)}{S(Q|Z)}\right\}^2
=cE\left\{\frac{f_0(T|Z)}{S_0(Q|Z)}-\frac{f(T|Z)}{S(Q|Z)}\right\}^2.
\end{align*}
That is
\begin{equation}\label{formula2}
E\left\{\frac{f_0(T|Z)}{S_0(Q|Z)}-\frac{f(T|Z)}{S(Q|Z)}\right\}^2 \ge cE\left\{\frac{S_0(T|Z)}{S_0(Q|Z)}-\frac{S(T|Z)}{S(Q|Z)}\right\}^2.
\end{equation}
And
\begin{align*}
E\left\{\frac{f_0(T|Z)}{S_0(Q|Z)}-\frac{f(T|Z)}{S(Q|Z)}\right\}^2
=E&\left\{\frac{S_0(T|Z)}{S_0(Q|Z)}e^{\beta_0Z}\lambda_0(T)-\frac{S(T|Z)}{S(Q|Z)}e^{\beta Z}\lambda(T)\right\}^2\\
=E&\left[\left\{\frac{S_0(T|Z)}{S_0(Q|Z)}-\frac{S(T|Z)}{S(Q|Z)}\right\}e^{\beta_0Z}\lambda_0(T)\right.\\
&\left.+\frac{S(T|Z)}{S(Q|Z)}\left(e^{\beta_0Z}\lambda_0(T)-e^{\beta Z}\lambda(T)\right)\right]^2.
\end{align*}

Let $A=\left\{\frac{S_0(T|Z)}{S_0(Q|Z)}-\frac{S(T|Z)}{S(Q|Z)}\right\}e^{\beta_0Z}\lambda_0(T)$ and $B=\frac{S(T|Z)}{S(Q|Z)}\left(e^{\beta_0Z}\lambda_0(T)-e^{\beta Z}\lambda(T)\right)$, then
$$E\left\{\frac{f_0(T|Z)}{S_0(Q|Z)}-\frac{f(T|Z)}{S(Q|Z)}\right\}^2=E(A+B)^2.$$

If $EA^2\ge cEB^2$ for a small $c>0$, since $$E\left\{\frac{S_0(T|Z)}{S_0(Q|Z)}-\frac{S(T|Z)}{S(Q|Z)}^2\right\}\ge cEA^2$$ and $$E\left(e^{\beta_0Z}\lambda_0(T)-e^{\beta Z}\lambda(T)\right)^2 \le cEB^2,$$ we have
$$E\left\{\frac{S_0(T|Z)}{S_0(Q|Z)}-\frac{S(T|Z)}{S(Q|Z)}^2\right\}\ge cE\left(e^{\beta_0Z}\lambda_0(T)-e^{\beta Z}\lambda(T)\right)^2$$
Also by (\ref{formula2})
$$E\left\{\frac{f_0(T|Z)}{S_0(Q|Z)}-\frac{f(T|Z)}{S(Q|Z)}\right\}^2 \ge cE\left(e^{\beta_0Z}\lambda_0(T)-e^{\beta Z}\lambda(T)\right)^2.$$
On the other hand, if $EA^2< cEB^2$ for small $c>0$, then by $2EAB<c\sqrt{EA^2 EB^2}<cEB^2$,
$$E\left\{\frac{f_0(T|Z)}{S_0(Q|Z)}-\frac{f(T|Z)}{S(Q|Z)}\right\}^2=EA^2+2EAB+EB^2>EB^2-cEB^2>cEB^2.$$
Then we still have
$$E\left\{\frac{f_0(T|Z)}{S_0(Q|Z)}-\frac{f(T|Z)}{S(Q|Z)}\right\}^2 \ge cE\left(e^{\beta_0Z}\lambda_0(T)-e^{\beta Z}\lambda(T)\right)^2.$$
Hence by (\ref{formula1}),
$$\mathbb{M}(\theta_0)-\mathbb{M}(\theta)\ge cP\left(L(\beta_0,\lambda_0)- L(\beta,\lambda)\right)^2\ge cE\left(e^{\beta_0Z}\lambda_0(T)-e^{\beta Z}\lambda(T)\right)^2.$$
By the same arguments on page 2126 and 2127 in \cite{wellner:zhang:07}, given C6

$$E\left(e^{\beta_0Z}\lambda_0(T)-e^{\beta Z}\lambda(T)\right)^2\ge c\left\{\|\beta-\beta_0|^2+\|\lambda-\lambda_0\|_{L_2(\nu)}^2\right\},$$  Hence,
$$\mathbb{M}(\theta_0)-\mathbb{M}(\theta)\ge c\left\{|\beta-\beta_0|^2+\|\lambda-\lambda_0\|_{L_2(\nu)}^2\right\}
=cd(\theta_0,\theta)^2~~\square.$$

\begin{lem}\label{lem:3}
Given C2. Then there exists $\lambda_n\in\mathfrak{F}_n$ such that $$\parallel \lambda_n-\lambda_0\parallel_\infty\le c(n^{-pv}).$$
\end{lem}

{\em\underline{Proof of Lemma \ref{lem:3}}}

Suppose the spline coefficients of $\lambda_n$ are chosen as $\alpha_j=\lambda_0(\eta_j)$ where $\eta_j$'s are defined by (0.5)
in proof of Lemma 0.2 in \cite{wu:zhang:12}.
Given C2, by Jackson's Theorem on page 149 in \cite{deboor:01} it is easy to see that
$\parallel \lambda_n-\lambda_0 \parallel_\infty \le c(n^{-p\kappa})$.

To complete the proof, it remains to show that $\alpha_j$'s satisfy the conditions for $\mathfrak{F}_n$.
\begin{enumerate}
\item
By $\alpha_j=\lambda_0(\eta_j)$, we have $a_0\le\alpha_j\le b_0<K\tau b_0$ for all $\alpha_j$'s.
\item
$\int_0^\tau\lambda_n(t)dt\le\int_0^\tau\max_{j}\alpha_jdt\le \tau b_0$.
\item
$\frac{\left|\alpha_{j+1}-\alpha_j\right|}{\max_j\Delta_j}
\le\frac{\left|\alpha_{j+1}-\alpha_j\right|}{\eta_{j+1}-\eta_j}
=\frac{\left|\lambda_0(\eta_{j+1})-\lambda_0(\eta_{j})\right|}{\eta_{j+1}-\eta_j}
\le\max_{t\in[0,\tau]}\left|\lambda'_0(t)\right|=d_0<Kd_0$.~~$\square$
\end{enumerate}

\begin{lem}\label{lem:4}
Given C5 and let $q$ and $t$ be any two fixed numbers with $0<q\le \tau_1$ and $q<t\le \tau$.  Then  $$\int_q^tf^2(x)dx\le c\left\|f\right\|_{L_2(\nu)}^2.$$
\end{lem}

{\em\underline{Proof of Lemma \ref{lem:4}}}

 C5 implies that for any $q$   with $0<q\le \tau_1$  the joint density of $(Q,T)$ has a positive lower bound in region
 $[0\le Q\le q, q\le T\le \tau]$.  Then for any $t$ with $q<t\le \tau$ we have
\begin{align*}
\int_q^tf^2(x)dx&\le c\int_0^q\int_q^\tau f^2(x)dxdy\le c\int_0^q\int_q^\tau f^2(x)d\nu(y,x)\\
&\le c\int_0^\tau \int_y^\tau f^2(x)d\nu(y,x)= c\left\|f\right\|_{L_2(\nu)}^2.~~\square
\end{align*}

\newpage


\end{document}